\documentclass[%
 reprint,
 amsmath,amssymb,
 aps,
]{revtex4-2}

\usepackage{graphicx}% Include figure files
\usepackage{dcolumn}% Align table columns on decimal point
\usepackage{bm}% bold math
\usepackage[mathlines]{lineno}% Enable numbering of text and display math
\usepackage{siunitx}
\usepackage[]{hyperref}

\begin{document}

\preprint{APS/123-QED}

\title{Constraints on the Axion-Photon Coupling Using Stellar Modelling}% Force line breaks with \\
%\thanks{A footnote to the article title}%

\author{David Fordham}
\email{david.fordham@tecnico.ulisboa.pt}
% \altaffiliation[Also at ]{Departamento de F\'{\i}sica, Instituto Superior T\'ecnico, Universidade de Lisboa, Lisboa, Portugal}%Lines break automatically or can be forced with \\
\author{Ilídio Lopes}%
% \email{Second.Author@institution.edu}
\affiliation{%
 Centro de Astrof\'{\i}sica e Gravita\c c\~ao  - CENTRA, Departamento de F\'{\i}sica, Instituto Superior T\'ecnico - IST, Universidade de Lisboa - UL, Av. Rovisco Pais 1, 1049-001 Lisboa, Portugal}%

%\date{\today}% It is always \today, today,
             %  but any date may be explicitly specified

\begin{abstract}
{Asteroseismology has been shown to be, together with stellar modelling, an invaluable tool in constraining properties of novel physics.}
  % aims heading (mandatory)
   {In this work, we study for the first time the influence of axionic production in the evolution of a late main-sequence star, comparing computational models with observational data in order to constrain the axion-photon $g_{a\gamma}$ coupling parameter.}
  % methods heading (mandatory)
   {We first perform a high-precision calibration of a stellar model to our target star, in order to obtain a benchmark for our other diagnostics. We then apply a two-stage test, first using global quantities and then resorting to precision seismic ratios.}
  % results heading (mandatory)
   {We find that seismology allows us to place an independent upper bound of $g_{a\gamma} \leq 0.98\times 10^{-10} $ GeV$^{-1}$ at a $68\%$ confidence level (CL), in the same order of magnitude as both the most recent constraints from the observation of globular clusters and previous bounds obtained through stellar modelling, but more stringent than most current direct axion detections. We also suggest a more conservative limit of $g_{a\gamma} \leq 1.38\times 10^{-10} $ GeV$^{-1}$ at a $95\%$ {CL}. Moreover, this new diagnostic method can be applied to stellar data that will be obtained in future asteroseismic projects.}\end{abstract}

%\keywords{Suggested keywords}%Use showkeys class option if keyword
                              %display desired
\maketitle

%\tableofcontents

\section{Introduction}

Axions currently stand out as one of the most promising candidates for dark matter \citep{Bertone:2016nfn,Chadha-Day:2021szb}. Originally predicted by Weinberg and Wilczek \citep{Weinberg:1977ma, Wilczek:1977pj}, their potential existence would not only offer insight into the nature of dark matter but also address a fundamental puzzle in modern physics -- the strong charge-parity (CP) problem. Several proposals have been put forward to explain the absence of observed CP violation in strong interactions, with the Peccei-Quinn theory \citep{Peccei:1977ur, Peccei:1977hh} emerging as one of the leading solutions. In this approach, a global chiral U(1) symmetry is imposed, allowing for the CP-violating phase term to vanish from the extended lagrangian of the standard model. This symmetry can be spontaneously broken, giving rise to a new particle that could solve both this problem and the dark matter hypotheses -- the axion. A large range of experiments have therefore been taking place in order to constrain the properties of the axion, specifically its mass $m_a$ and the strength of its coupling to the photon $g_{a\gamma}$. 

Axion helioscopes search for axions produced in the interior of the Sun. One of the leading prototypes is the CERN Axion Solar Telescope -- CAST \citep{CAST:2017uph} -- that makes use of a dipole magnet to produce an adjustable magnetic field that allows for the conversion of axions to X-rays through the Primakoff effect \citep{Primakoff:1951iae}, easily detectable by using a focusing mirror system for X-rays and appropriate detectors. Recent developments report a sensitivity to the axion-photon coupling constant of $g_{a\gamma}\leq \SI{0.88e-10}{\per\giga\electronvolt}$ at $95\%$ CL for $m_a \leq \SI{0.02}{\electronvolt}/c^2$, which constitutes our best direct constraint to date for very light axion-like particles. For realistic quantum chromodynamics (QCD) axions, the bound lies at $g_{a\gamma} < \SI{2.3e-10}{\per\giga\electronvolt}$ at $95\%$ CL.
    ``Light shining through walls'' experiments are also widely used in the search effort, where laser photons are shot through a magnetic field to be converted into axions, directed towards an opaque wall and finally reconverted into photons in a second magnetic field. The leading experiment of this type is the ALPS-II experiment, which started in 2019 \citep{Spector:2016vwo}. 
    Axion haloscopes function in a similar way to helioscopes. However, they do not restrict their search to solar axions, rather functioning with the assumption of much less energetic particles, detecting the conversion of axions into microwaves in a microwave cavity. The leading current axion haloscope project is the Axion Dark Matter Experiment -- ADMX -- having its most recent instalment started in 2016 \citep{ADMX:2020ote}.
    
In addition to the already mentioned direct detection methods, stars have also been used as far-away laboratories for probing axionic limits and dark matter theories in general, by taking advantage of effects such as energy loss streams caused by the creation of a new particle. The strongest current bound on this axion-photon coupling constant actually comes from the study of globular cluster stars, that allow for precision testing of stellar-evolution theory \citep{Ayala:2014pea}. The Primakoff effect would increase the core energy release of stars on the horizontal branch, decreasing the lifetime of this stage of evolution by a certain amount. By counting the number of red giants compared to that of horizontal branch stars of various globular clusters, a conservative bound was found at $g_{a\gamma}\leq \SI{6.6e-11}{\per\giga\electronvolt}$ for a $95\%$ CL.

Asteroseismology has also been taken advantage of as a way to probe new physics with stars \citep[e.g.][]{Casanellas:2010he}. In this work, we use for the first time high precision measurements of stellar oscillations in order to probe the production of axions inside late main-sequence (MS) stars. This is done by analysing the oscillation frequencies of stars, that provide us invaluable information about their structure. As mentioned before, the insertion of the axion in our paradigm introduces a new energy transport mechanism, that can lead to changes in the stellar interior, which we will also study in this article. These changes can affect the stars' internal oscillations, often in a more noticeable way than its global spectroscopic quantities. The inner workings of the star can then be studied using relevant asteroseismic diagnostics, rendering this field a powerful precision probing tool. Missions such as CoRoT \citep{Auvergne:2009tq} and Kepler \citep{Kepler:2010xwo} have been invaluable in obtaining the oscillation frequencies of a myriad of stars located at diverse points of the stellar evolution track, opening the door to using seismology on stars other than the Sun. Having this range of stars available greatly increases the effectiveness of asteroseismic studies, as we can choose exactly the kind of ``laboratory'' that suits our experiments, depending on the characteristics of each star and its evolution stage. While its main goal will be discovering habitable extra-solar planets, the PLAnetary Transits and Oscillations of stars (PLATO) \citep{plato} mission will provide us with high-precision measurements of both spectroscopic and seismic stellar quantities for a vastly greater number of stars, more than 200,000 cool dwarfs and subgiants (SG) \citep{2010ASPC..430..260C} over those currently observed by Kepler. This will allow for a precise calibration of stellar models and a broader study of the impact axions have on stars through asteroseismology. As shown in this work, high-precision seismic observations of solar-like stars can play an essential role to this end. 

In the following section, we describe the relevant axionic interactions in the stellar interior. In section \ref{sec::seismo}, we address the basis of asteroseismology, important quantities for describing stellar oscillations, and the seismic diagnostic actually used. Then in section \ref{sec::cal} we focus on the calibration method used, and the diagnostic used to ensure the quality of our models. We go on to calibrate a late MS star in section \ref{sec::models}, without the presence of axions, choosing the best no-axion model as a benchmark model based on which we evolve a series of axion models that we then compare to each other. After that, in section \ref{sec::analysis}, we take the models evolved in section \ref{sec::models} and perform a second diagnostic, this time a precision seismic one, to observe with extra sensitivity the effect the axion has in the stellar interior, and to obtain limits on axion properties. Finally, we present conclusions and closing remarks in the last section.

\section{On Axions and their Interactions with the Stellar Interior} \label{sec::interactions}

Allowing for a solution of two of the most pressing issues in modern physics, axionic theories have been developed at length since the prediction of this particle in 1978. Originally created to specifically tackle the strong {CP} problem \citep{ParticleDataGroup:2014cgo}, if discovered, this particle would provide a validation for the Peccei-Quinn mechanism, protecting the strong interaction from {CP}-violating effects \citep{Peccei:1977ur,Peccei:1977hh}, as suggested by the experimental evidence of the absence of a neutron electric dipole moment.

The issue stems from the existence of a {CP}-violating term in the {QCD} Lagrangian, $ \mathcal{L}_\Theta = - \overline{\Theta} (\alpha_s/8\pi) G^{\mu\nu a} \tilde{G}_{\mu\nu}^a$.
 
  Here, $-\pi \leq \overline{\Theta} \leq +\pi$ is the effective $\Theta$ parameter after diagonalising quark masses, and $G_{\mu\nu}^a$ is the colour field strength tensor. Experimental constraints on the neutron electric dipole moment \citep{Burghoff:2011xk} imply an extremely low value of $ | \overline{\Theta} | \lessapprox 10^{-10}$, which has no theoretical reasoning. 
  
 At energies below the electroweak scale, the global Peccei-Quinn symmetry $\mathrm{U(1)_{PQ}}$ would solve this problem when its associated current features an $\mathrm{SU(3)_C\times U(1)}$ chiral anomaly, as it can be spontaneously broken due to the axion's non-zero triangle coupling to gluons \citep{Ringwald:2015lqa},
  
  \begin{equation} \label{eq::CP_term}
  \mathcal{L} = \left(\frac{\phi_a}{f_a} -\overline{\Theta} \right)\frac{\alpha_s}{8\pi}G^{\mu\nu a} \tilde{G}_{\mu\nu}^a,
  \end{equation}

\noindent with $\phi_a$ being the axion field and $f_a$ the axion decay constant. Non-perturbative fluctuations of the gluon fields in the {QCD} framework induce a potential for $\phi_a$, its minimum lying at $\phi_a= \overline{\Theta} f_a$, thereby cancelling out the {CP}-violating term, as seen in equation \eqref{eq::CP_term}. This axion is therefore a pseudo Nambu-Goldstone boson, as it gets a small mass due to the same non-perturbative effects \citep{Ringwald:2015lqa}.

After extensive laboratory and astrophysical evidence, the ``standard axion'' of Weinberg and Wilczek has been excluded in favour of an alternative, ``invisible'' particle \citep{Raffelt:1985nk}.
In particular, the Kim, Shifman, Vainshtein and Zakharov (KSVZ) axion \citep{Kim:1979if,Shifman:1979if}, which we will focus on in this study, couples to regular matter mainly through a double photon vertex, and in stellar interiors its production mechanism is largely assured by photon conversion through the Primakoff effect \citep{Primakoff:1951iae}. This effect allows for the photoproduction of axions in the presence of electrons and nuclei,

\begin{equation}
    \gamma + (e^-,Ze) \rightarrow (e^-,Ze)+a.
\end{equation}

The effective coupling of axions to a photon pair can be described by the lagrangian density \citep[e.g.][]{Raffelt:1998fy},

\begin{equation}
\mathcal{L}_{a\gamma} = -\frac{g_{a\gamma}}{4}F_{\mu\nu}\tilde{F}^{\mu\nu}\phi_a   =g_{a\gamma}\phi_a\mathbf{E}\cdot \mathbf{B},
\end{equation}
	
\noindent where $F$ is the electromagnetic field-strength tensor, $a$ is the axion field and $g_{a\gamma}$ is a model-dependant coupling constant of dimension (energy)$^{-1}$, that is written as

\begin{equation}
g_{a\gamma} = \frac{\alpha}{2\pi f_a}\left(\frac{E}{N} - 1.92(4)\right) = \left(0.20(3)\frac{E}{N}-0.39(1)\right)\frac{m_a}{\mathrm{GeV}^2},
\end{equation}

\noindent where $E$ and $N$ are the electromagnetic and colour anomalies of the axial current, and the $1.92(4)$ is a result of the mixing of the axion with the QCD mesons below the confinement scale \citep{Li:2020naa}. For the {KSVZ} model studied in this work \cite{Kim:1979if,Shifman:1979if}, $E/N=0$, although a vast array of $E/N$ values is possible in different theoretical models.

This coupling constant is constructed in such a way that it depends linearly on the mass of the axion, and since this particular kind of axion has a high decay constant it becomes very difficult to detect -- invisible -- making $g_{a\gamma}$ one of the key factors in understanding the relevance of this model. 

In the interior of stars, most of the axionic energy loss is due to this Primakoff conversion, through photon-nucleus scattering that is mediated by virtual photons of the electrostatic potential of the nucleus. After taking into account the Debye-Huckel effect \citep{Huckel1924}, the resulting expression for the energy loss rate in a non-degenerate medium is well established as being \citep{Raffelt:1998fy}

\begin{equation}
\varepsilon_{a} = \frac{g_{a\gamma} ^2 T^7}{4\pi ^2 \rho} \xi^2 f(\xi^2),
\end{equation}

\noindent where the function $f$ is defined as an integral over the photon distribution \citep{Raffelt:1990yz} written as a function of $\xi \equiv \frac{\hbar c k_S}{2k_B T}$, with $\hbar$ the reduced Planck constant, $c$ the speed of light in vacuum, $k_B$ the Boltzmann constant, and $k_S$ the Debye-Huckel screening wavenumber, given by \citet{Huckel1924}, $k_S^2 \equiv 4\pi \alpha \left(\frac{\hbar c}{k_B T}\right) \sum_{i = e, \mathrm{ions}} n_i Z_i ^2$, with $\alpha$ the fine-structure constant, $Z_i$ the atomic number, and $n_i$ the ion or electron number density. 

This expression can be rewritten in a simpler form as

\begin{equation}\label{eq::axion_lum}
    \varepsilon_{a}=283.16 \times g_{10}^{2} T_{8}^{7} \rho_{3}^{-1} \xi^{2}f(\xi^{2}) \; \mathrm{erg} / \mathrm{g} / \mathrm{s},
\end{equation}

\noindent where $g_{10}\equiv g_{a\gamma}/(10^{-10}\, \textrm{GeV}^{-1})$, $\rho_3 \equiv \rho/(10^3\, \textrm{g/cm}^3)$, and $T_{8}\equiv T/(10^{8}\, \textrm{K})$. The numerical factor has been rectified as seen in the appendix section of \citet{Choplin:2017auq}. 

The $f(\xi^2)$ distribution must also be parametrised in a simple yet accurate manner, in order to include it in a stellar evolution code. We proceed to use the approximation proposed by \citet{Friedland:2013fse}, which captures the distribution's limits and its intermediate regime with an accuracy of over $98\%$ across the entire range of $\xi$,

\begin{equation}
	f(\xi^2) \approx \left( \frac{1.037}{1.01+\xi^2/5.4} + \frac{1.037}{44+0.628\xi^2}\right) \ln\left(3.85+\frac{3.99}{\xi^2}\right).
\end{equation}

Typical values of this function are, for example, $\xi^2\sim 12$ and $\xi^2f(\xi^2)\sim 6$ for the Sun, and $\xi^2\sim 2.5$ and $\xi^2f(\xi^2)\sim 3$ for low-mass He burning stars \citep{Raffelt:1990yz}.

It has been shown that axions have a great effect in the helium burning phases of stellar evolution \citep{Friedland:2013fse, Choplin:2017auq}, for central temperatures between $10^8$ K and $4\times 10^8$ K. In these stages it is possible to probe axionic losses purely through an analysis of spectral parameters. However, we have very little data on oscillation frequencies for stars at this evolutionary stage, and modelling them produces wildly different evolutionary tracks depending on the stellar code that is used \citep{Agrawal:2020znh}. Luckily, this cooling effect also occurs in the MS and SG branches, 
for which we have a broad catalogue and precise seismic and photometric measurements. The scale of the effect is however much smaller than that observed during the helium burning stages, calling for a more sensitive field such as asteroseismology in order to apply competitive constraints.

\section{Asteroseismology as a Probing Tool} \label{sec::seismo}

\subsection{How stars oscillate}

Asteroseismology has been proven to be an indispensable tool to probe the stellar interior through the study of its oscillations, and has been used a number of times to constrain the parameters of different kinds of dark matter particles \citep[e.g.][]{Casanellas:2010he, Rato:2021tfc}. In this field of study, we assume spherical symmetry, which means that oscillations can then be described as caused by a combination of standing waves characterised by radial $n$, spherical $\ell$ and azimuthal $m$ numbers. The most prevalent kinds of standing waves considered are acoustic p--modes and gravity g--modes \citep[e.g.][]{dalsgaard}. In addition to these, mixed modes can arise in stars that have evolved off the MS, having a p-mode character in convective regions and a g-mode character in radiative regions. 

P--modes, which have pressure as their restoring force, are stronger in the envelope of MS and SG stars. The most common diagnostics to retrieve information from these modes is the so-called large frequency separation, i.e., the difference in frequency between subsequent modes with the same angular degree \citep{dalsgaard}:

\begin{equation}
    \Delta \nu_{n,l} = \nu_{n,\, l}-\nu_{n-1,\, l} \approx  \left(2\int_0^R \frac{dr}{c(r)} \right)^{-1},
\end{equation}

\noindent where $c(r)$ is the speed of sound at radius $r$ and $R$ is the total radius of the star.

We can also define a small frequency separation, highly sensitive to thermodynamic conditions in the stellar core \citep{1980ApJS...43..469T},

\begin{equation}
    \delta \nu_{n,l} = \nu_{n,\, l}-\nu_{n-1,\, l+2}. 
\end{equation}

On the other hand we have g--modes, or gravity modes, that have buoyancy as their restoring force, and are especially sensitive to the inner core of stars. They are often described by the separation in period $\Delta \Pi_\ell$ \citep{dalsgaard},

\begin{equation}\label{eq:delta_pi}
\Delta \Pi_\ell = \frac{2\pi^2}{\sqrt{\ell(\ell+1)}}\left(\int_{r_1}^{r_2}N\frac{dr}{r}\right)^{-1} \equiv \frac{\Pi_0}{\sqrt{\ell(\ell+1)}},
\end{equation}

\noindent where $N$ is the Brunt-Väisälä (or buoyancy) frequency, and $r_1$ and $r_2$ are the turning points of the g-mode cavity, thus signalling that $\Delta \Pi_\ell$ is directly related to the size of the convective core.

These two types of waves propagate very differently. On the one hand, acoustic waves have a lower propagation cavity bound related to the characteristic acoustic frequency, being observed predominately in the outer layers of the star. On the other hand, gravity waves are bound by the buoyancy frequency, that is directly related to the size of the radiative core of the star. In the convective zones we would have $N^2<0$, and so the g-modes would be evanescent in these regions.

\subsection{Seismic ratio diagnostic}

Although stellar models can make robust predictions on global spectroscopic observations, many times they do not mirror the workings of the stellar interior. It is here that asteroseismology comes in as a high precision diagnostic tool, using oscillation analysis as a way to select the best models. 

The seismic ratios of small to large frequency separations have been shown to be excellent diagnostics of the interior of solar-like stars \citep[e.g.][]{Casanellas:2010he,refId1}. Since we do not have observed modes with $\ell > 2$, we decide to use the $r_{02}$ quantity,
\begin{equation}
    r_{02}(n) = \frac{  \delta \nu_{n,0}}{\Delta \nu_{n,1}}.
\end{equation}

{As individual oscillation frequencies are highly sensitive to near surface effects, this ratio is built in a way that cancels out the latter, leaving us with a quantity that is extremely sensitive to the stellar interior \citep{refId0}. This allows us to probe the region of the star where axion production is expected to be the strongest -- the core -- with a simple ratio, through the direct comparison of its observed value to those obtained through stellar models. }

It is important to note that this is a highly sensitive diagnostic, which is best used as a final phase of a two-step method. This creates a robust rejection process, where the first and coarser step is applied through the calibration of stellar models to global observable parameters of the star.

\section{Calibration Procedure}\label{sec::cal}

\subsection{Selecting the Star}

In terms of an ideal candidate, we are looking for a star with precisely determined spectroscopic and asteroseismic properties. This means that it should have a relatively high number of detected oscillation modes, with their respective frequencies, and coherently modelled by previous sources. In terms of evolution stage, we search for a late MS star, in order to detect the maximum amount of axion cooling effect while still having access to a large number of well-fitted p-modes. Stars with a mass close to that of the Sun present themselves as attractive study cases, both because they are easier to detect, since more massive stars deplete their resource faster and are therefore shorted lived, and because the luminosity related to axionic emission represents a larger fraction of the total energy loss of these stars, hopefully evidencing clearer axion signatures. Moreover, the internal physics of low-mass stars at this evolution stage is known at a higher detail than that of more massive and evolved targets, whose modelling is rarely even congruent between stellar evolution codes \citep{Agrawal:2020znh}. This is the case since the physics of these stars is closer to that of the Sun, and due to the large number of observations of such targets. 
We do not choose the Sun itself as we want to establish a diagnostic that is not reliant on helioseismology, but rather on the kind of data resulting from missions such as {Kepler} \citep{Kepler:2010xwo}. The method can thus be applied to a statistically significant number of targets, and benefit from improved data delivered by efforts such as the upcoming PLATO \citep{plato}. 

As for the oscillation classification, we search for a simple star according to \citet{Appourchaux:2012zm}'s classification in order to obtain clearly identifiable oscillation modes that can be reliably utilised in precision diagnostics, as other kinds of stars present avoided crossings which lead to deviations from the regular frequency spacing and void the validity of certain seismological tests.

{We thus select {KIC} 6933899, a late G0.5IV MS star \citep{Molenda-Zakowicz:2013waa} with a previously modelled mass of between 1.10 and 1.14 M$_\odot$ \citep{Mathur:2012sk} and 33 detected oscillation modes with a precision in the order of $\SI{0.1}{\micro\hertz}$ \citep{Appourchaux:2012zm}. This target has an acoustic behaviour, as it exhibits mostly simple p-modes, 
 rendering diagnostics such as the $r_{02}$ ratio appealing to probe the physics of its core.}

 \subsection{Obtaining a Benchmark Model}

For the purpose of stellar modelling, we resort to the Modules for Experiments in Stellar Astrophysics ({MESA}) \citep{Paxton2011, Paxton2013, Paxton2015, Paxton2018, Paxton2019}, an open-source 1-D stellar evolution code that allows the user to obtain models for a wide array of stellar objects given a set of input parameters.
 
The MESA EOS is a blend of the OPAL \citep{Rogers2002}, SCVH \citep{Saumon1995}, FreeEOS \citep{Irwin2004}, HELM \citep{Timmes2000}, PC \citep{Potekhin2010}, and Skye \citep{Jermyn2021} EOSes. Radiative opacities are primarily from OPAL \citep{Iglesias1993,Iglesias1996}, with low-temperature data from \citet{Ferguson2005} and the high-temperature, Compton-scattering dominated regime by \citet{Poutanen2017}. Electron conduction opacities are from \citet{Cassisi2007}. Nuclear reaction rates are from JINA REACLIB \citep{Cyburt2010}, NACRE \citep{Angulo1999} and additional tabulated weak reaction rates \citet{Fuller1985, Oda1994,Langanke2000}. Screening is included via the prescription of \citet{Chugunov2007}. Thermal neutrino loss rates are from \citet{Itoh1996}.

 The convection theory utilised is the mixing-length theory \citep{1968pss..book.....C}, and radiative levitation is neglected. The metallicity [FeH] is calculated by $\mathrm{[FeH]} = \log[(Z/X)/(Z/X)_\odot]$, where the solar reference $(Z/X)_\odot = 0.02293$ is computed by \citet{Bahcall:2005va} based on the solar metal mixture of \citet{Grevesse:1998bj}.

We use the \emph{astero} module available in {MESA} in order to obtain a high quality stellar model calibrated to our target star. Using this tool, we produce a series of evolutionary models using \{M, [FeH], Y, $\alpha_{\mathrm{MLT}}$,$f_{ov}$\}, respectively initial mass, metallicity, helium abundance, mixing length coefficient and overshooting parameter, as initial input parameters. We then produce a stellar model that evolves from a chemically-homogeneous pre main-sequence to current age, being subjected to a direct comparison with real data at this step. The quality of each model can be assessed by calculating the default quantity $\chi^2_{star} = \frac{1}{3}\chi^2_{spec} + \frac{2}{3}\chi^2_{seis}$ \citep{Paxton2013,Metcalfe:2012mu,Rato:2021tfc}, which is a composition with default weights of the quadratic deviation of the spectral and seismic quantities,

\begin{equation}\label{chi2}
     \chi^2_{spec/seis} = \frac{1}{N} \sum_{i=1}^{N} \left( \frac{X_i^{mod} - X_i^{obs}}{\sigma_{X_i^{obs}}}\right)^2,
\end{equation}

\noindent where $N$ is the number of parameters, $X_i^{mod}$ and $X_i^{obs}$ are the stellar model and observed values of the i$^{\mathrm{th}}$ parameter, respectively, with $\sigma_{X_i^{obs}}$ being the observational uncertainty. The set of observational spectral parameters that we use are $\{T_{eff},\, \log g,\, [Fe/H]\} = \{5830\pm 70\,\mathrm{K},\, 4.02\pm 0.08\, \mathrm{cm/s}^2,\, 0.01\pm 0.07 \} $, respectively the effective temperature, logarithm of surface gravity, and metallicity, and the observational seismic parameter set used is $\{\Delta \nu\} = \{72.15\pm 0.25 \, \mu\mathrm{Hz}\}$, the large frequency separation, as taken from \citet{Mathur:2012sk}.

As for the optimisation procedure, we resort to the Nelder-Mead method, as applied throughout literature \citep[e.g.][]{Capelo:2020lha}, which uses a direct search downhill simplex algorithm \citep{Nelder:1965zz}. It consists in minimising the $\chi^2_{star}$ by varying the input parameters and finding an optimal set that produces a stellar model with a group of output parameters \{$T_{eff}$, $\log g$, [Fe/H], $\Delta \nu$\} as close to the observable ones as possible. After the number of optimisation steps is achieved or the $\chi^2_{star}$ values stagnate, we can conclude that the algorithm has converged. A number of measures can be taken to avoid the detection of a local minimum instead of the global one, such as a preliminary parameter scan resorting to a rough grid search. We choose to use only the global $\Delta \nu$ parameter for the seismic input parameters instead of the oscillation frequencies so as to compute our models in an achievable computational time, as well as to be able to use the frequencies in a posterior phase of a double-stage diagnostic, in our case by analysing the $r_{02}$ ratios.

\begin{table*}
    \centering
    \caption{Resulting calibrated models for {KIC} 6933899. Columns with symbols not mentioned before include: $\tau$ -- age, R -- total radius, L -- luminosity, $\rho_c$ -- central density. }
    \begin{tabular}{cccccccccccccc}
    \hline \hline
     g$_{10}$ & M/M$_{\odot}$ & $\tau$(Gyr) & R/R$_{\odot}$ & L/L$_{\odot}$ &
T$_{\mathrm{eff}}$(K) & log(g/[cm/s$^2$])&f$_{ov}$& Y & [FeH] &
$\alpha_{\mathrm{MLT}}$ & $\rho_c$(g/cm$^3$)& $\Delta\nu$($\mu$Hz) & $\chi_{star}^2$($10^{-2}$)\\ \hline
  0.0&   1.14 & 8.20 & 1.67 & 2.97 & 5862&4.051 & 0.004 & 0.248 & 0.004 & 2.092 &1803& 72.10  & 1.18\\ 
  1.0&1.12&8.51&1.66&2.97&5877&4.048&0.005&0.252&0.006&1.999&1949&72.10&1.12\\
  1.1* & 1.12&8.24&1.66&2.93&5860&4.048&0.004&0.256&0.019&2.030&1941&72.10&0.01\\
  2.0&1.10&8.54&1.64&2.84&5832&4.045&0.003&0.258&0.009&1.989&2070&72.15&1.08\\
  3.0&1.08&8.55&1.64&2.81&5829&4.043&0.006&0.263&0.011&2.056&2494&72.15&0.90\\  \hline
  \multicolumn{12}{l}{\footnotesize{*The g$_{10}$ parameter was kept free for this calibration.}}\\
    \end{tabular}
    \label{coisaout}
\end{table*}

We proceed as described and obtain a Benchmark Model ({BM}) for which main output parameters resulting from this calibration can be seen in table \ref{coisaout}. Comparing these results to those obtained by \citet{Mathur:2012sk} with their three grid-based results, all fall within $3\sigma$ of our model. The radius stands out as the quantity that is, although statistically coherent, systematically lower than our value, which is expected for grid-based methods \citep{Mathur:2012sk}. We therefore consider the calibration a satisfactory {BM}.

\begin{figure}
     \centering

         \includegraphics[width=\columnwidth]{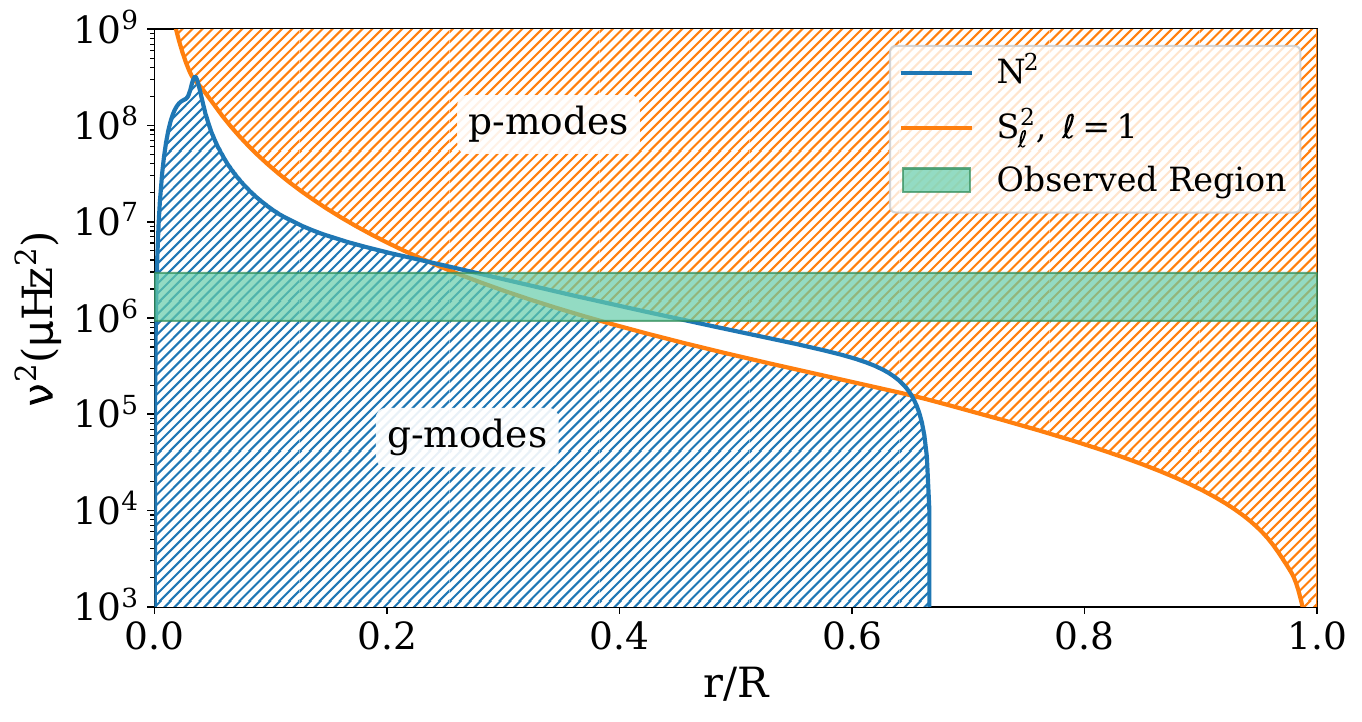}
 
     \caption{Propagation diagram of {KIC} 6933899. The g-- and p--mode propagation cavities are represented respectively by the blue and orange areas. Modes observed by the {Kepler} mission \citep{Kepler:2010xwo} reside within the green region.}
     \label{fig:mode_trap}
\end{figure}

Our models show a radiative core and convective envelope, and, as expected for stars at this evolution stage for this mass, a helium core engulfed in a hydrogen shell. The propagation diagram of this star is shown in figure \ref{fig:mode_trap}, along with the frequency range of its observed modes, and is typical to that of a late MS star \citep[e.g.][]{hekker_mazumdar_2013}.

\section{Stellar Models} \label{sec::models}

\subsection{Comparison of Fully Calibrated Axion Models}

In order to study how axions affect stellar evolution, we take advantage of the fact that MESA allows us to treat the axion-photon coupling constant $g_{a\gamma}$ as a free calibration parameter. This can be achieved by introducing the axion emission effect in the form of equation \ref{eq::axion_lum} into an extra MESA subroutine that modifies the default settings. In the case of the axion cooling patch, it builds on the non-nuclear neutrino losses routine, adding the energy stream generated by the Primakoff effect to the non-nuclear energy sources calculated at each iteration. These subroutines are baked into the default ones during the control and startup phases, and allow us to alter or add to the standard physics with no need to rewrite the base code. This way, the code can safely run as usual, ensuring solutions for the equations of stellar structure that guarantee a hydrostatic equilibrium at each step, with the new physics enclosed in the appropriate loops \citep{Paxton2011, Paxton2013, Paxton2015, Paxton2018, Paxton2019}. 

For a number of fixed $g_{10}$ values, we produce another set of full calibrations, with the input parameters as those used for the benchmark model. We also include one calibration where $g_{10}$ was kept free. Once again, each calibration generates a number of models, and for each one we choose the model with the lowest $\chi_{star}^2$. The results can be seen in table \ref{coisaout}. Although most of these macroscopic parameters remain relatively similar, which is expected due to the low impact of axion cooling in MS and SG stars, there are a few trends that can be observed. Focusing first on the models where $g_{10}$ was not kept free, the most noticeable parameter trend is the increase of the stellar age with $g_{10}$, by up to $4\%$. Given the introduction of a novel outflow of energy from the stellar interior, this is in agreement with our expectations, as more losses lead to faster burning \citep{Friedland:2013fse}. Furthermore, a characteristic of late MS and early SG stars is the presence of an expanding envelope, as the radius increases to balance the radiation pressure derived from internal nuclear reactions \citep{Kippenhahn:2012qhp}. It is therefore striking to see a decreasing trend in the stellar radius, despite the increasing ages. This can, however, be once again explained due to the axion losses coming from the stellar interior, that ``cool down'' the core and diminish the relative outwards pressure, causing the radius to in fact decrease, which in turn leads to a lower luminosity. This cooling is evident for higher $g_{10}$ values by looking directly at the calibrated effective temperature, which keeps these calibration results consistent. Finally, the decreasing trend in the stellar mass might be occurring so as to keep increasingly older stellar models at the same evolutionary standpoint, indirectly leading to a steady decrease in surface gravity, and the increase in central density as a direct result of a cooler stellar core. It is also important to state that these trends all coexist as responses both to the introduction of this energy stream and themselves, as the models try to adjust to the observable parameters.

Regarding the calibration with $g_{10}$ as a free parameter, we now observe a departure from a few of the previously stated trends. However, this is likely a case of overfitting, where we have too many free parameters for the problem we want to calibrate, as evidenced by the strikingly low $\chi^2_{star}$. Another difference that leads us to disregard this method of calibration is the extraordinarily high initial metallicity, up by $375\%$ compared to the BM. This once again supports the two-phased constraining method that has been applied in this work.

\begin{figure}
     \centering
 
    \includegraphics[width=\columnwidth]{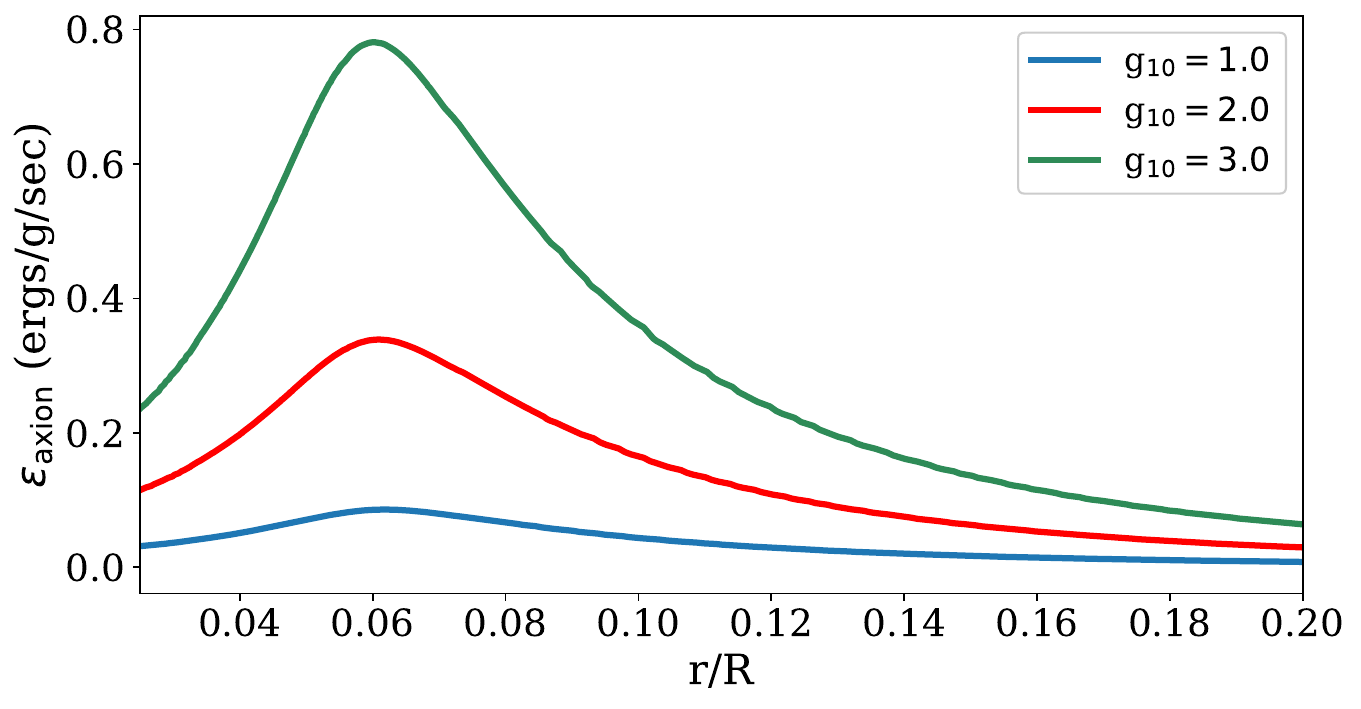}
  
     \caption{Energy loss rate profile in {KIC} 6933899's stellar interior caused by the Primakoff effect, eq. \eqref{eq::axion_lum}, for various $g_{10}$ values.}
     \label{fig:axlum}
\end{figure}

\begin{figure}[t]
     \centering

         \includegraphics[width=\columnwidth]{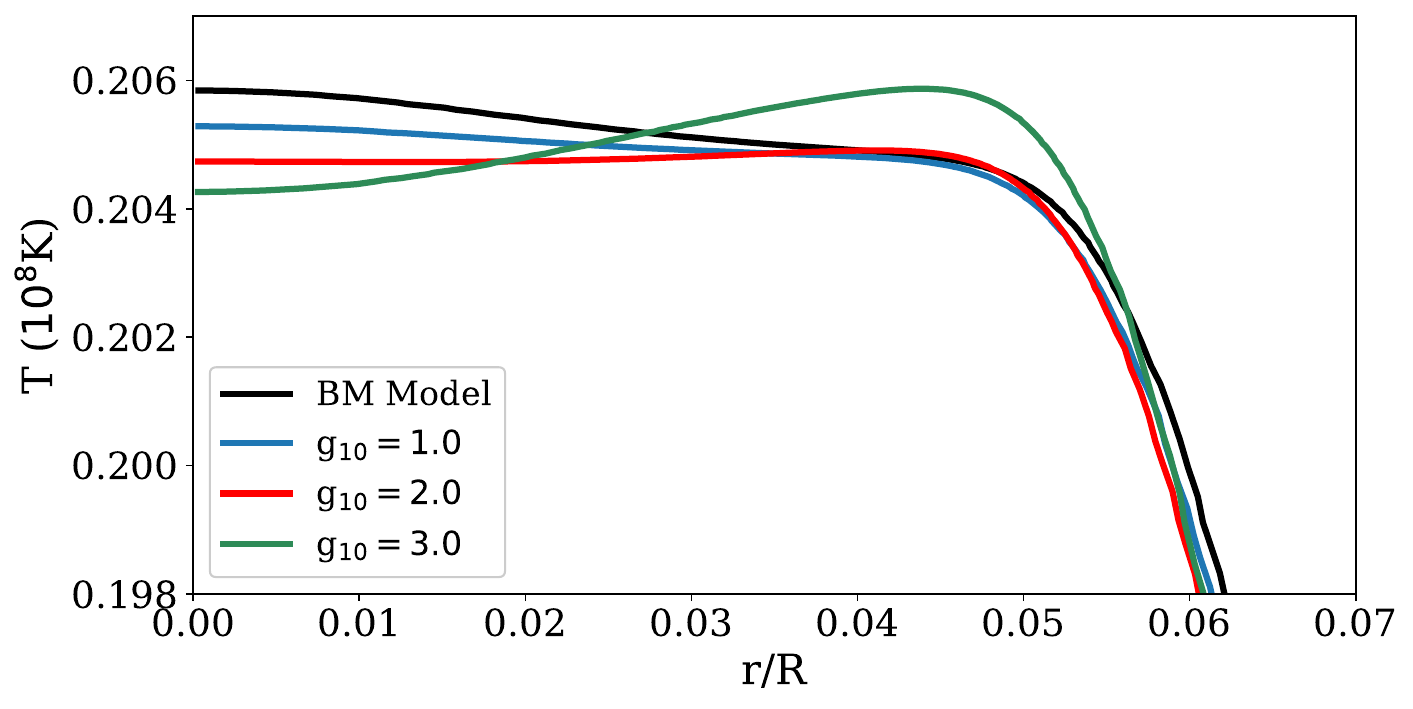}
 
     \caption{Temperature profile of {KIC} 6933899's core for various $g_{10}$ values and the {BM}.}
     \label{fig:temperature}
\end{figure}

To verify whether the Primakoff effect results in actual measurable differences in the stellar interior, we now plot and analyse a few key profiles of the stellar interior for these different coupling values, $g_{10} = 0.0, 1.0, 2.0, 3.0$.

We start by directly plotting the energy production profile generated by axion emission in the star, which can be seen in figure \ref{fig:axlum}. Here we can instantly verify that this particle is indeed produced mostly near the centre of the star -- between 0.050 and 0.075 of the total radius of each model. This is also where the spike of energy produced through pp-chain reactions can be observed in this star, near the burning shell that engulfs the inner helium core, a characteristic of late main-sequence and subgiant stars. The insertion of this axion model creates an energy loss profile that, at its most intense point, produces a channel with an energy loss comparable to $0.1\%$ to $1.0\%$ of that released by nuclear reactions, from lowest to highest coupling value. This instantly tells us, for example, that $g_{10}=3.0$ is too strong of a coupling parameter, as it has been shown that axionic energy production happens mostly during the helium-burning phases of evolution \citep[e.g.][]{Friedland:2013fse}, so it should not come close to compete with that of nuclear reactions during the late MS stage. We also verify that, at this evolution phase, the axionic energy channel is $10^4$ to $10^5$ times more efficient in evacuating energy than the neutrino loss stream.

The temperature profile of our star is displayed in figure \ref{fig:temperature}. It is reassuring to verify that a higher coupling value cools the stellar core, leaving the rest of its profile virtually unaffected, as is predicted for the axion cooling phenomenon enabled by the Primakoff effect. We also confirm that the core temperature does seem to vary an appreciable amount, up to a $1\%$ decrease, indicating actual change in the stellar structure, but not enough to render this an effective constraining method without resorting to a precision tool such as asteroseismology. Furthermore, a trend seems to be observed where slightly higher coupling values promote the existence of an isothermal core. For greater $g_{10}$'s, however, the profile becomes profoundly altered, with the temperature actually rising throughout the core. This is another indicator that some coupling strengths can actually improve the modelling process, as stars near the terminal-age main sequence are known to display isothermal cores \citep{Kippenhahn:2012qhp}, due to the inert helium core that is developed through the exhaustion of hydrogen in the nucleus \citep{10.1088/978-0-7503-1278-3ch14}. Other coupling values can then be considered less suitable candidates, although not completely rejected, as would be the case, for example, for $g_{10}=3.0$, which completely destroys the isothermal profile of the stellar core. This would indicate that the innermost core is in fact not inert for this kind of star, which is not the established consensus. Finally, a temperature spike can be observed around the same region where axion production is strongest, as a way to balance the decrease in internal temperatures generated by the axion losses.

\begin{figure}[t]
     \centering

         \includegraphics[width=\columnwidth]{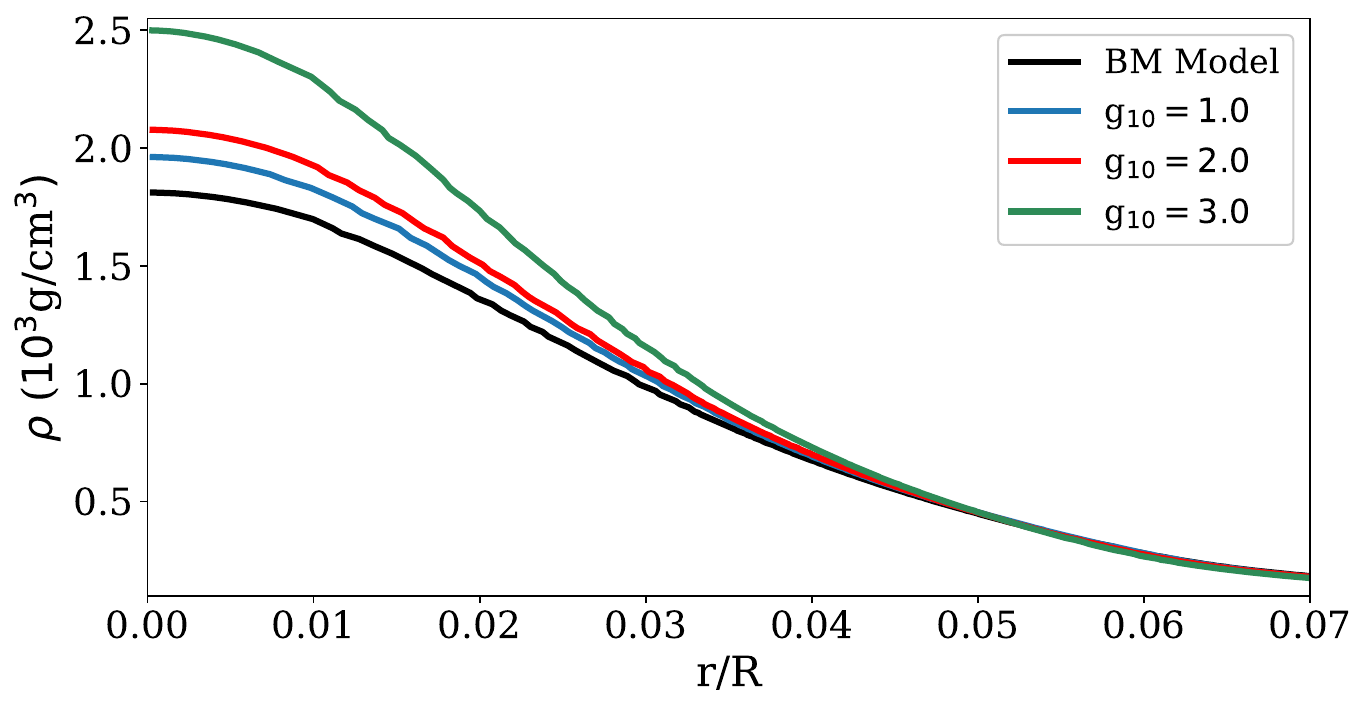}
 
     \caption{Baryonic density profile of {KIC} 6933899's core for axionic models and the {BM}.}
     \label{fig:dens}
\end{figure}

At a first glance, looking at figure \ref{fig:temperature} and equation \eqref{eq::axion_lum}, one might expect to see the peak of the axionic emission right at the centre of the core, and not around the burning shell, as seen in figure \ref{fig:axlum}.
Although the production rate depends on the seventh power of the temperature, it is indeed the whole factor of $T_{8}^{7} \rho_{3}^{-1}$ that determines the shape of the emission profile. By plotting the baryonic density profile in the appropriate units of eq. \eqref{eq::axion_lum}, as displayed in figure \ref{fig:dens}, we can see its sharp decrease throughout the core.
This variation is much more significant than the change in temperature for the same range, even when considering a power of seven, leading to a rise in the axionic emissions throughout the core until the hydrogen shell is reached.

\subsection{First Phase $\chi^2$ Analysis}

After this preliminary analysis, we now want to fine-scan the $g_{a\gamma}$ parameter space. To achieve this, we fix the model's initial parameters with those of the BM, and simply evolve a sequence of stellar models whose only varying parameter is the axion-photon coupling constant $g_{a\gamma}$. This method allows us to rapidly produce a large set of stellar models, covering all the parameter space of interest. We confirm the validity of this approximation by fully calibrating a second model with a coupling parameter in the order of the highest values considered, thus verifying that the optimised global parameters do not vary substantially around those of the {BM}. For example, optimal masses all range between 1.10 and 1.15 M$_\odot$, and ages between 8.0 and 8.6 Gyr. To compare our pallet of models, we conveniently present some of our $\chi^2$ tests in a normalised manner,

\begin{equation}
    \overline{\chi}^2_x =\frac{\chi^2_x}{\chi^2_{BMx} },
\end{equation}

\noindent which just means that a $\chi^2$ test applied to any quantity $x$ will be normalised by the same value obtained for the benchmark model.

\begin{figure}[t]
     \centering

         \includegraphics[width=\columnwidth]{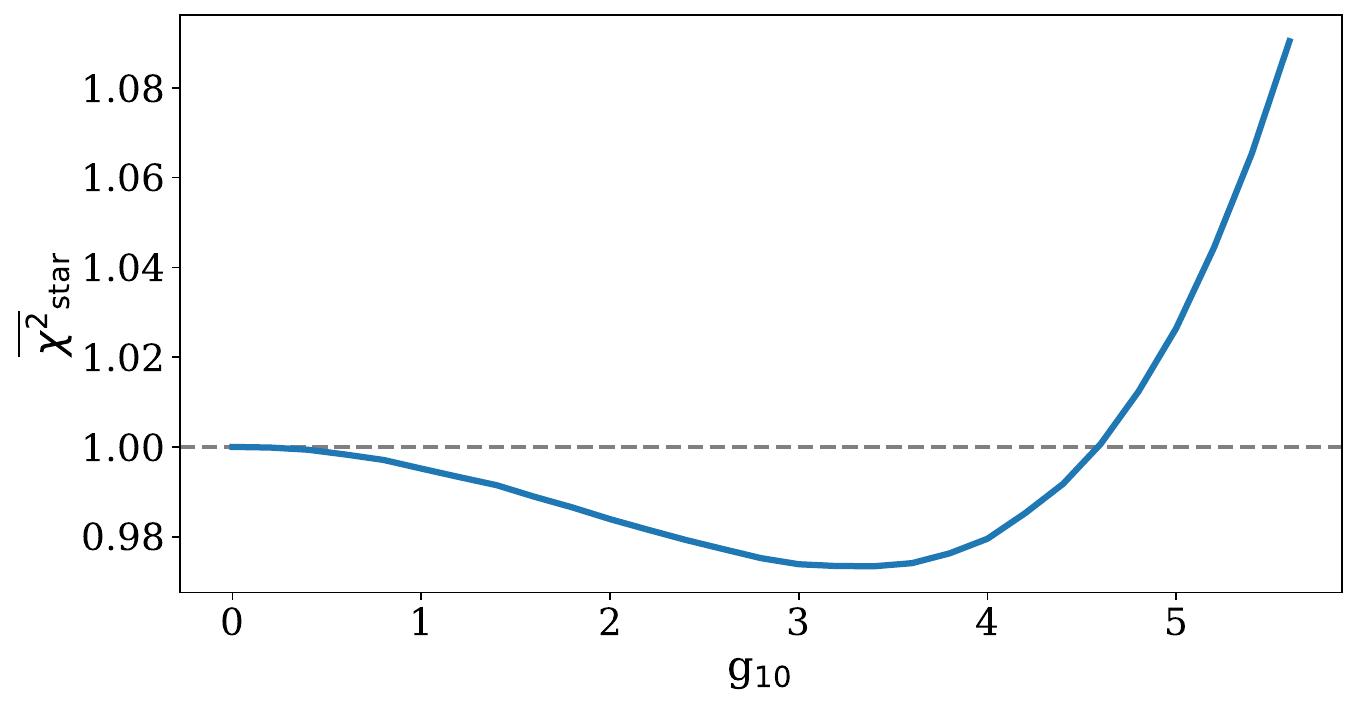}
      
     \caption{Result of $\overline{\chi}^2_{star}$ test for each coupling value. Points below $\overline{\chi}^2_{star} = 1$ indicate couplings that lead to a better adjustment to observable parameters than the benchmark model. All models with $g_{10} \geq 4.6$ are outperformed by the benchmark model ({BM}).}
     \label{fig:g10}
\end{figure}

Our efforts result in figure \ref{fig:g10}, where we can see how the $\chi^2_{star}$ changes for each $g_{10}$. We can observe two regions, one where $\overline{\chi}^2_{star} $ is less than 1, and another where it is greater than 1. Our star presents a $\chi^2$ minimum at a non-zero coupling value, $g_{10} = 3.4$. This suggests plausibility in the use of this axion model, as there is a continuous range of $g_{10}$ consistent with current bounds that result in models that better reproduce observable parameters. This area is followed by a sharp rise in $\chi^2$, signalling a hard upper bound on $g_{10}$, as this range of values would cause the stellar models to diverge too much from the observed quantities.

Based on this first analysis, we impose a preliminary upper bound on the coupling parameter by excluding all models from which the inclusion of the axion cooling effect for a higher coupling parameter would result exclusively in models worse than the one with no axion effect. This gives us a preliminary bound of $g_{10} \leq 4.5$. 

We choose to do this instead of picking the $g_{10}$ that results in the minimum value of $\overline{\chi}^2_{star}$ as being the optimal axion-photon coupling value because fitting procedures based solely on global parameters can contain a non-negligible degree of correlation. Furthermore, the overall change in $\chi^2$ is quite small, never surpassing $1\%$, which further proves the need for a second more sensitive diagnostic in this procedure. However, this bound is not final, and serves purely as a starting point for the second phase of this exclusion method, since these $\chi^2$ variations are not statistically significant enough to impose definite constraints.

\begin{figure}[t]
         \centering
         \includegraphics[width=\columnwidth]{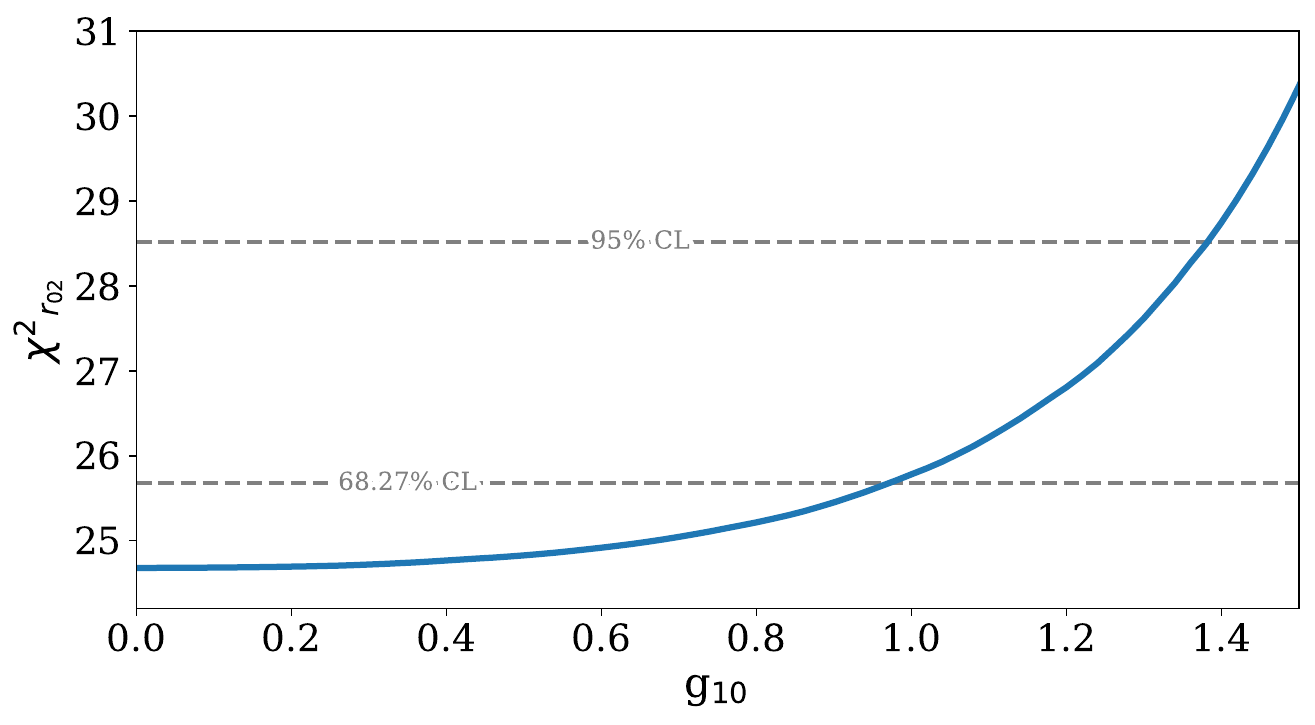}
     \caption{Result of the reduced ${\chi}^2_{r_{02}}$ statistic for the $10^{-11}$ GeV$^{-1}$ coupling scale, as well as the confidence levels at $68\%$ and $95\%$ with a $dof=7-1$.}
     
     \label{fig:chir02_close}
\end{figure}

\section{Asteroseismic Analysis}\label{sec::analysis}

Using the stellar models computed for the grid of $g_{10}$ values for the result presented in figure \ref{fig:g10}, we calculate the eigenmodes of each one of them using {GYRE} \citep{Townsend:2013lua} and match the radial order of each observed oscillation mode \citep{Mathur:2012sk,Appourchaux:2012zm} to that of a modelled one. This is achieved by comparing frequency values $\nu$ and spherical orders $\ell$.

We now delve into the structure of our seismic diagnosis, and we calculate the ratio $r_{02}(n)$ for our observed and modelled modes. To obtain a quantitative measure of the quality of each range of ratios, we introduce the quantity

\begin{equation}\label{chir02}
    \chi^2_{r_{02}} =  \sum_{n=15}^{21} \left( \frac{r_{02}^{mod} (n) - r_{02}^{obs} (n)}{\sigma_{r_{02}^{obs}(n)}}\right)^2, 
\end{equation}

\noindent for all observed mode pairings that allow for the calculation of this ratio, in our case from $n=15 $ to $n=21$. Given the higher sensitivity of seismic ratios, we choose to apply a step of 0.01 for $g_{10}$ in the grid search we perform.

\begin{figure}
         \centering
         \includegraphics[width=\columnwidth]{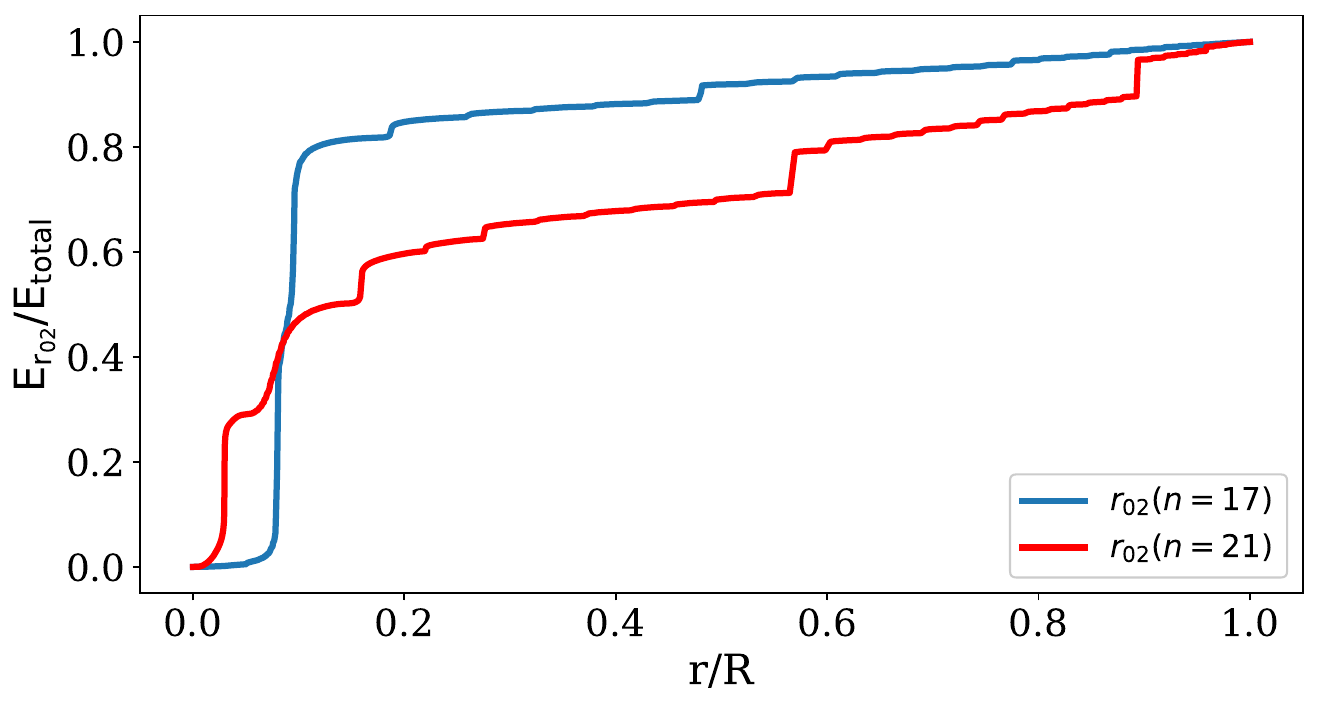}
     \caption{Profile of the fraction of total inertia carried by the $r_{02}$ mode combinations for the lowest and highest radial order used in our ratio calculations. }
     \label{fig:E_frac}
\end{figure}
    
The result can be seen in figure \ref{fig:chir02_close}. This time, instead of normalised to the benchmark value, we choose to show the reduced $\chi^2$, with $dof=7-1$ degrees of freedom. As expected, this closeup allows us to understand the changes occurring inside this star with a newfound precision. {By tracing the lines referent to a $68\%$ and $95\%$ {CL}s for a $\chi^2$ distribution with the BM value as its base}, we are now able to state that the Primakoff effect with an axion-photon coupling constant of $g_{10}\leq 0.97$ is compatible with the modelling of {KIC} 6933899 at a $68\%$ {CL}, which is in accord with previous limits proposed through stellar modelling by \citet{Friedland:2013fse}, though utilising a better known stellar evolutionary phase and applying a precision diagnostic through asteroseismology.

A more conservative bound can be obtained by looking at the $95\%$ {CL} line, which lands at $g_{10}\leq 1.38$, also in line with our current understanding of the {KSVZ} axion model. Furthermore, profiles such as those shown in figures \ref{fig:axlum} and \ref{fig:temperature} reassure us that this order of magnitude of $g_{a\gamma}$ reflects indeed actual change in the stellar interior. It is also an improvement on most recent {CAST} findings \citep{CAST:2017uph}, which placed a bound of $g_{10}< 2.3$ for realistic QCD axions, indicating that it might be productive to improve the sensitivity of observational experiments. In fact, this is exactly what the International Axion Observatory ({IAXO}) \citep{iaxo}, {CAST}'s follow-up, will do, allowing for the observation of masses between 1 meV to 1 eV. It is scheduled to be launched in 2028.

In order to check how effective $r_{02}$ is at probing the seismic stellar core, we will begin by retrieving the differential inertia $dE_{n,l}/dr$, as defined in \citet{dalsgaard}, of each mode used in our diagnostic. We then calculate the resulting differential inertia profile for the combination of modes used in $r_{02}$, and integrate it throughout the stellar radius, as a way to see wherein the contributions to the total inertia of this ratio lie, i.e

\begin{equation}
    E_{r_{02}}(n,r) \equiv   \int_0^{r}\left|\frac{\frac{dE_{n, 0}}{dr}(r)-\frac{dE_{n-1, 2}}{dr}(r)}{\frac{dE_{n, 1}}{dr}(r)-\frac{dE_{n-1, 1}}{dr}(r)}\right|dr.
\end{equation}

We sketch this profile for the no-axion model using the lowest and highest $n$ considered in our calculations, normalising it by the total inertia of each combination of modes $E_{\mathrm{total}}(n) \equiv E_{r_{02}}(n,R)$. The result can bee seen in figure \ref{fig:E_frac}.

We can now easily verify that, although the considered eigenmodes acquire almost all of their inertia in the stellar envelope, most of that of $r_{02}$ originates in the stellar interior. Despite showing a few small bumps in the outer layers of the star, 80$\%$ of the total $r_{02}$ inertia for lower radial orders {is generated at approximately $r/R<0.2$, while the value becomes $60\%$  for the higher $n$, providing us with an insight to just how sensitive this seismic ratio really is to the stellar core, and confirming its reliability for the chosen solar-like star}. 

\section{Conclusion}

We have explored for the first time the impact of the axion cooling effect on the solar-like target {KIC} 6933899. For this purpose, we calibrated a benchmark stellar model with a null axion-photon coupling, that resulted in global parameters compatible with previous modelisations. We then introduced an axionic energy loss stream and created new stellar models with an increasing $g_{a\gamma}$ strength, comparing how each model relates to observable parameters relatively to the BM. 

We found that axions are produced mostly in the stellar core, which is in agreement with other sources \citep[e.g.][]{Raffelt:1990yz}. In this late MS star, the peak axion production is located in the vicinity of the burning hydrogen shell, where most of the production of nuclear energy takes place. Furthermore, stellar models that include axion emission tend towards a more isothermal core, in line with what is expected for a late MS star, until the $g_{a\gamma}$ coupling becomes too strong to the point where its internal temperature profile would be completely altered.

Through this first global comparison, we were able to place a preliminary bound of around $g_{a\gamma} \leq 4.6\times 10^{-10} $ GeV$^{-1}$. This value is quite larger than that of the most competitive limits, and can be explained due to the relatively small contribution of the Primakoff effect in this evolution stage, which results in relatively small changes of the global parameters. This reinforces the importance of applying a second stage to the diagnostic. By taking advantage of the large amount of precise seismic data that is available, we can implement a much more sensitive and statistically significant analysis.

Finally, we used the ratio $r_{02}$ as an asteroseismic diagnostic to probe the stellar interior with greater precision. We found that the sensitivity of this method would be optimal for scanning the $10^{-11}$ GeV$^{-1}$ order of $g_{a\gamma}$, a degree of magnitude below that of the previous diagnostic. We arrive at the limit of  $g_{a\gamma} \leq 0.97\times 10^{-10} $ GeV$^{-1}$, at a $68\%$ {CL}, and of $g_{a\gamma} \leq 1.38\times 10^{-10} $ GeV$^{-1}$, at a $95\%$ {CL}.

At $68\%$ {CL}, this constraint is on par with that obtained by \citet{Ayala:2014pea} at the same confidence level, which was calculated by counting the number of horizontal branch stars and of red giants in globular clusters. It is also in line with that obtained by \citet{Friedland:2012hj}, through the modelling of massive red-giants but without resorting to precision seismology or stellar calibration. 

It is worth noting that these three methods use different sets of astronomical data and focus on distinct stages of stellar evolution, which highlights the capability of stellar astrophysics to constrain the axion-photon coupling constant, while being consistent with laboratory searches. In particular, an asteroseismic study of a solar-like star is a very powerful tool for this kind of diagnostic, since their internal physics is quite well-known compared to stars at later evolution stages. This allows us to study oscillation modes detected with a \SI{0.1}{\micro\hertz} precision, resulting in competitive constraints obtained in a reliable manner.

This bound is also stricter than that obtained by the {CAST} experiment \citep{CAST:2017uph} for realistic QCD axions, providing a big expectation for the results of its follow-up project -- {IAXO} \citep{iaxo}.

%\quad

%%%%%%%%%%%%%%%%%%%%%%%%%%%%%%%%%%%%%%%%%%%%%
%\acknowledgments
\section*{Acknowledgments}
%\medskip\noindent

DF and IL thank Bill Paxton and all MESA contributors for making their code and work publicly available, and Alexander Friedland and Maurizio Giannotti for their contribution with the axion cooling patch. We thank Funda\c c\~ao para a Ci\^encia e Tecnologia (FCT), Portugal, for the financial support to the Center for Astrophysics and Gravitation (CENTRA/IST/ULisboa) through the Grant Project~No.~UIDB/00099/2020  and Grant No. PTDC/FIS-AST/28920/2017.  

\newpage

\bibliography{bibi}% Produces the bibliography via BibTeX.

%apsrev4-2.bst 2019-01-14 (MD) hand-edited version of apsrev4-1.bst
%Control: key (0)
%Control: author (8) initials jnrlst
%Control: editor formatted (1) identically to author
%Control: production of article title (0) allowed
%Control: page (0) single
%Control: year (1) truncated
%Control: production of eprint (0) enabled
\begin{thebibliography}{72}%
\makeatletter
\providecommand \@ifxundefined [1]{%
 \@ifx{#1\undefined}
}%
\providecommand \@ifnum [1]{%
 \ifnum #1\expandafter \@firstoftwo
 \else \expandafter \@secondoftwo
 \fi
}%
\providecommand \@ifx [1]{%
 \ifx #1\expandafter \@firstoftwo
 \else \expandafter \@secondoftwo
 \fi
}%
\providecommand \natexlab [1]{#1}%
\providecommand \enquote  [1]{``#1''}%
\providecommand \bibnamefont  [1]{#1}%
\providecommand \bibfnamefont [1]{#1}%
\providecommand \citenamefont [1]{#1}%
\providecommand \href@noop [0]{\@secondoftwo}%
\providecommand \href [0]{\begingroup \@sanitize@url \@href}%
\providecommand \@href[1]{\@@startlink{#1}\@@href}%
\providecommand \@@href[1]{\endgroup#1\@@endlink}%
\providecommand \@sanitize@url [0]{\catcode `\\12\catcode `\$12\catcode
  `\&12\catcode `\#12\catcode `\^12\catcode `\_12\catcode `\%12\relax}%
\providecommand \@@startlink[1]{}%
\providecommand \@@endlink[0]{}%
\providecommand \url  [0]{\begingroup\@sanitize@url \@url }%
\providecommand \@url [1]{\endgroup\@href {#1}{\urlprefix }}%
\providecommand \urlprefix  [0]{URL }%
\providecommand \Eprint [0]{\href }%
\providecommand \doibase [0]{https://doi.org/}%
\providecommand \selectlanguage [0]{\@gobble}%
\providecommand \bibinfo  [0]{\@secondoftwo}%
\providecommand \bibfield  [0]{\@secondoftwo}%
\providecommand \translation [1]{[#1]}%
\providecommand \BibitemOpen [0]{}%
\providecommand \bibitemStop [0]{}%
\providecommand \bibitemNoStop [0]{.\EOS\space}%
\providecommand \EOS [0]{\spacefactor3000\relax}%
\providecommand \BibitemShut  [1]{\csname bibitem#1\endcsname}%
\let\auto@bib@innerbib\@empty
%</preamble>
\bibitem [{\citenamefont {Bertone}\ and\ \citenamefont
  {Hooper}(2018)}]{Bertone:2016nfn}%
  \BibitemOpen
  \bibfield  {author} {\bibinfo {author} {\bibfnamefont {G.}~\bibnamefont
  {Bertone}}\ and\ \bibinfo {author} {\bibfnamefont {D.}~\bibnamefont
  {Hooper}},\ }\bibfield  {title} {\bibinfo {title} {{History of dark
  matter}},\ }\href {https://doi.org/10.1103/RevModPhys.90.045002} {\bibfield
  {journal} {\bibinfo  {journal} {Rev. Mod. Phys.}\ }\textbf {\bibinfo {volume}
  {90}},\ \bibinfo {pages} {045002} (\bibinfo {year} {2018})},\ \Eprint
  {https://arxiv.org/abs/1605.04909} {arXiv:1605.04909 [astro-ph.CO]}
  \BibitemShut {NoStop}%
\bibitem [{\citenamefont {Chadha-Day}\ \emph {et~al.}(2022)\citenamefont
  {Chadha-Day}, \citenamefont {Ellis},\ and\ \citenamefont
  {Marsh}}]{Chadha-Day:2021szb}%
  \BibitemOpen
  \bibfield  {author} {\bibinfo {author} {\bibfnamefont {F.}~\bibnamefont
  {Chadha-Day}}, \bibinfo {author} {\bibfnamefont {J.}~\bibnamefont {Ellis}},\
  and\ \bibinfo {author} {\bibfnamefont {D.~J.~E.}\ \bibnamefont {Marsh}},\
  }\bibfield  {title} {\bibinfo {title} {{Axion dark matter: What is it and why
  now?}},\ }\href {https://doi.org/10.1126/sciadv.abj3618} {\bibfield
  {journal} {\bibinfo  {journal} {Sci. Adv.}\ }\textbf {\bibinfo {volume}
  {8}},\ \bibinfo {pages} {abj3618} (\bibinfo {year} {2022})},\ \Eprint
  {https://arxiv.org/abs/2105.01406} {arXiv:2105.01406 [hep-ph]} \BibitemShut
  {NoStop}%
\bibitem [{\citenamefont {Weinberg}(1978)}]{Weinberg:1977ma}%
  \BibitemOpen
  \bibfield  {author} {\bibinfo {author} {\bibfnamefont {S.}~\bibnamefont
  {Weinberg}},\ }\bibfield  {title} {\bibinfo {title} {{A New Light Boson?}},\
  }\href {https://doi.org/10.1103/PhysRevLett.40.223} {\bibfield  {journal}
  {\bibinfo  {journal} {Phys. Rev. Lett.}\ }\textbf {\bibinfo {volume} {40}},\
  \bibinfo {pages} {223} (\bibinfo {year} {1978})}\BibitemShut {NoStop}%
\bibitem [{\citenamefont {Wilczek}(1978)}]{Wilczek:1977pj}%
  \BibitemOpen
  \bibfield  {author} {\bibinfo {author} {\bibfnamefont {F.}~\bibnamefont
  {Wilczek}},\ }\bibfield  {title} {\bibinfo {title} {{Problem of Strong $P$
  and $T$ Invariance in the Presence of Instantons}},\ }\href
  {https://doi.org/10.1103/PhysRevLett.40.279} {\bibfield  {journal} {\bibinfo
  {journal} {Phys. Rev. Lett.}\ }\textbf {\bibinfo {volume} {40}},\ \bibinfo
  {pages} {279} (\bibinfo {year} {1978})}\BibitemShut {NoStop}%
\bibitem [{\citenamefont {Peccei}\ and\ \citenamefont
  {Quinn}(1977{\natexlab{a}})}]{Peccei:1977ur}%
  \BibitemOpen
  \bibfield  {author} {\bibinfo {author} {\bibfnamefont {R.~D.}\ \bibnamefont
  {Peccei}}\ and\ \bibinfo {author} {\bibfnamefont {H.~R.}\ \bibnamefont
  {Quinn}},\ }\bibfield  {title} {\bibinfo {title} {{Constraints Imposed by CP
  Conservation in the Presence of Instantons}},\ }\href
  {https://doi.org/10.1103/PhysRevD.16.1791} {\bibfield  {journal} {\bibinfo
  {journal} {Phys. Rev. D}\ }\textbf {\bibinfo {volume} {16}},\ \bibinfo
  {pages} {1791} (\bibinfo {year} {1977}{\natexlab{a}})}\BibitemShut {NoStop}%
\bibitem [{\citenamefont {Peccei}\ and\ \citenamefont
  {Quinn}(1977{\natexlab{b}})}]{Peccei:1977hh}%
  \BibitemOpen
  \bibfield  {author} {\bibinfo {author} {\bibfnamefont {R.~D.}\ \bibnamefont
  {Peccei}}\ and\ \bibinfo {author} {\bibfnamefont {H.~R.}\ \bibnamefont
  {Quinn}},\ }\bibfield  {title} {\bibinfo {title} {{CP Conservation in the
  Presence of Instantons}},\ }\href
  {https://doi.org/10.1103/PhysRevLett.38.1440} {\bibfield  {journal} {\bibinfo
   {journal} {Phys. Rev. Lett.}\ }\textbf {\bibinfo {volume} {38}},\ \bibinfo
  {pages} {1440} (\bibinfo {year} {1977}{\natexlab{b}})}\BibitemShut {NoStop}%
\bibitem [{\citenamefont {Anastassopoulos}\ \emph {et~al.}(2017)\citenamefont
  {Anastassopoulos} \emph {et~al.}}]{CAST:2017uph}%
  \BibitemOpen
  \bibfield  {author} {\bibinfo {author} {\bibfnamefont {V.}~\bibnamefont
  {Anastassopoulos}} \emph {et~al.} (\bibinfo {collaboration} {CAST}),\
  }\bibfield  {title} {\bibinfo {title} {{New CAST Limit on the Axion-Photon
  Interaction}},\ }\href {https://doi.org/10.1038/nphys4109} {\bibfield
  {journal} {\bibinfo  {journal} {Nature Phys.}\ }\textbf {\bibinfo {volume}
  {13}},\ \bibinfo {pages} {584} (\bibinfo {year} {2017})},\ \Eprint
  {https://arxiv.org/abs/1705.02290} {arXiv:1705.02290 [hep-ex]} \BibitemShut
  {NoStop}%
\bibitem [{\citenamefont {Primakoff}(1951)}]{Primakoff:1951iae}%
  \BibitemOpen
  \bibfield  {author} {\bibinfo {author} {\bibfnamefont {H.}~\bibnamefont
  {Primakoff}},\ }\bibfield  {title} {\bibinfo {title} {{Photoproduction of
  neutral mesons in nuclear electric fields and the mean life of the neutral
  meson}},\ }\href {https://doi.org/10.1103/PhysRev.81.899} {\bibfield
  {journal} {\bibinfo  {journal} {Phys. Rev.}\ }\textbf {\bibinfo {volume}
  {81}},\ \bibinfo {pages} {899} (\bibinfo {year} {1951})}\BibitemShut
  {NoStop}%
\bibitem [{\citenamefont {Spector}(2017)}]{Spector:2016vwo}%
  \BibitemOpen
  \bibfield  {author} {\bibinfo {author} {\bibfnamefont {A.}~\bibnamefont
  {Spector}} (\bibinfo {collaboration} {ALPS}),\ }\bibfield  {title} {\bibinfo
  {title} {{ALPS II technical overview and status report}},\ }in\ \href
  {https://doi.org/10.3204/DESY-PROC-2009-03/Spector_Aaron} {\emph {\bibinfo
  {booktitle} {{12th Patras Workshop on Axions, WIMPs and WISPs}}}}\ (\bibinfo
  {year} {2017})\ pp.\ \bibinfo {pages} {133--136},\ \Eprint
  {https://arxiv.org/abs/1611.05863} {arXiv:1611.05863 [physics.ins-det]}
  \BibitemShut {NoStop}%
\bibitem [{\citenamefont {Khatiwada}\ \emph {et~al.}(2021)\citenamefont
  {Khatiwada} \emph {et~al.}}]{ADMX:2020ote}%
  \BibitemOpen
  \bibfield  {author} {\bibinfo {author} {\bibfnamefont {R.}~\bibnamefont
  {Khatiwada}} \emph {et~al.} (\bibinfo {collaboration} {ADMX}),\ }\bibfield
  {title} {\bibinfo {title} {{Axion Dark Matter Experiment: Detailed design~and
  operations}},\ }\href {https://doi.org/10.1063/5.0037857} {\bibfield
  {journal} {\bibinfo  {journal} {Rev. Sci. Instrum.}\ }\textbf {\bibinfo
  {volume} {92}},\ \bibinfo {pages} {124502} (\bibinfo {year} {2021})},\
  \Eprint {https://arxiv.org/abs/2010.00169} {arXiv:2010.00169 [astro-ph.IM]}
  \BibitemShut {NoStop}%
\bibitem [{\citenamefont {Ayala}\ \emph {et~al.}(2014)\citenamefont {Ayala},
  \citenamefont {Dom\'\i{}nguez}, \citenamefont {Giannotti}, \citenamefont
  {Mirizzi},\ and\ \citenamefont {Straniero}}]{Ayala:2014pea}%
  \BibitemOpen
  \bibfield  {author} {\bibinfo {author} {\bibfnamefont {A.}~\bibnamefont
  {Ayala}}, \bibinfo {author} {\bibfnamefont {I.}~\bibnamefont
  {Dom\'\i{}nguez}}, \bibinfo {author} {\bibfnamefont {M.}~\bibnamefont
  {Giannotti}}, \bibinfo {author} {\bibfnamefont {A.}~\bibnamefont {Mirizzi}},\
  and\ \bibinfo {author} {\bibfnamefont {O.}~\bibnamefont {Straniero}},\
  }\bibfield  {title} {\bibinfo {title} {{Revisiting the bound on axion-photon
  coupling from Globular Clusters}},\ }\href
  {https://doi.org/10.1103/PhysRevLett.113.191302} {\bibfield  {journal}
  {\bibinfo  {journal} {Phys. Rev. Lett.}\ }\textbf {\bibinfo {volume} {113}},\
  \bibinfo {pages} {191302} (\bibinfo {year} {2014})},\ \Eprint
  {https://arxiv.org/abs/1406.6053} {arXiv:1406.6053 [astro-ph.SR]}
  \BibitemShut {NoStop}%
\bibitem [{\citenamefont {Casanellas}\ and\ \citenamefont
  {Lopes}(2011)}]{Casanellas:2010he}%
  \BibitemOpen
  \bibfield  {author} {\bibinfo {author} {\bibfnamefont {J.}~\bibnamefont
  {Casanellas}}\ and\ \bibinfo {author} {\bibfnamefont {I.}~\bibnamefont
  {Lopes}},\ }\bibfield  {title} {\bibinfo {title} {{Towards the use of
  asteroseismology to investigate the nature of dark matter}},\ }\href
  {https://doi.org/10.1111/j.1365-2966.2010.17463.x} {\bibfield  {journal}
  {\bibinfo  {journal} {Mon. Not. Roy. Astron. Soc.}\ }\textbf {\bibinfo
  {volume} {410}},\ \bibinfo {pages} {535} (\bibinfo {year} {2011})},\ \Eprint
  {https://arxiv.org/abs/1008.0646} {arXiv:1008.0646 [astro-ph.CO]}
  \BibitemShut {NoStop}%
\bibitem [{\citenamefont {Auvergne}\ \emph {et~al.}(2009)\citenamefont
  {Auvergne}, \citenamefont {Bodin}, \citenamefont {Boisnard}, \citenamefont
  {Buey},\ and\ \citenamefont {Chaintreuil}}]{Auvergne:2009tq}%
  \BibitemOpen
  \bibfield  {author} {\bibinfo {author} {\bibfnamefont {M.}~\bibnamefont
  {Auvergne}}, \bibinfo {author} {\bibfnamefont {P.}~\bibnamefont {Bodin}},
  \bibinfo {author} {\bibfnamefont {L.}~\bibnamefont {Boisnard}}, \bibinfo
  {author} {\bibfnamefont {J.~T.}\ \bibnamefont {Buey}},\ and\ \bibinfo
  {author} {\bibfnamefont {S.}~\bibnamefont {Chaintreuil}} (\bibinfo
  {collaboration} {CoRoT Team}),\ }\bibfield  {title} {\bibinfo {title} {{The
  CoRoT satellite in flight : description and performance}},\ }\href
  {https://doi.org/10.1051/0004-6361/200810860} {\bibfield  {journal} {\bibinfo
   {journal} {Astron. Astrophys.}\ }\textbf {\bibinfo {volume} {506}},\
  \bibinfo {pages} {411} (\bibinfo {year} {2009})},\ \Eprint
  {https://arxiv.org/abs/0901.2206} {arXiv:0901.2206 [astro-ph.SR]}
  \BibitemShut {NoStop}%
\bibitem [{\citenamefont {Borucki}\ \emph {et~al.}(2010)\citenamefont {Borucki}
  \emph {et~al.}}]{Kepler:2010xwo}%
  \BibitemOpen
  \bibfield  {author} {\bibinfo {author} {\bibfnamefont {W.~J.}\ \bibnamefont
  {Borucki}} \emph {et~al.} (\bibinfo {collaboration} {Kepler}),\ }\bibfield
  {title} {\bibinfo {title} {{Kepler Planet-Detection Mission: Introduction and
  First Results}},\ }\href {https://doi.org/10.1126/science.1185402} {\bibfield
   {journal} {\bibinfo  {journal} {Science}\ }\textbf {\bibinfo {volume}
  {327}},\ \bibinfo {pages} {977} (\bibinfo {year} {2010})}\BibitemShut
  {NoStop}%
\bibitem [{\citenamefont {Rauer}\ \emph {et~al.}(2014)\citenamefont {Rauer},
  \citenamefont {Catala}, \citenamefont {Aerts}, \citenamefont {Appourchaux},
  \citenamefont {Benz}, \citenamefont {Brandeker}, \citenamefont
  {Christensen-Dalsgaard}, \citenamefont {Deleuil}, \citenamefont {Gizon},
  \citenamefont {Goupil}, \citenamefont {G{\"u}del}, \citenamefont
  {Janot-Pacheco}, \citenamefont {Mas-Hesse}, \citenamefont {Pagano},
  \citenamefont {Piotto}, \citenamefont {Pollacco}, \citenamefont {Santos},
  \citenamefont {Smith}, \citenamefont {Su{\'a}rez}, \citenamefont {Szab{\'o}},
  \citenamefont {Udry}, \citenamefont {Adibekyan}, \citenamefont {Alibert},
  \citenamefont {Almenara}, \citenamefont {Amaro-Seoane}, \citenamefont {Eiff},
  \citenamefont {Asplund}, \citenamefont {Antonello}, \citenamefont {Barnes},
  \citenamefont {Baudin}, \citenamefont {Belkacem}, \citenamefont {Bergemann},
  \citenamefont {Bihain}, \citenamefont {Birch}, \citenamefont {Bonfils},
  \citenamefont {Boisse}, \citenamefont {Bonomo}, \citenamefont {Borsa},
  \citenamefont {Brand{\~a}o}, \citenamefont {Brocato}, \citenamefont {Brun},
  \citenamefont {Burleigh}, \citenamefont {Burston}, \citenamefont {Cabrera},
  \citenamefont {Cassisi}, \citenamefont {Chaplin}, \citenamefont {Charpinet},
  \citenamefont {Chiappini}, \citenamefont {Church}, \citenamefont {Csizmadia},
  \citenamefont {Cunha}, \citenamefont {Damasso}, \citenamefont {Davies},
  \citenamefont {Deeg}, \citenamefont {D{\'\i}az}, \citenamefont {Dreizler},
  \citenamefont {Dreyer}, \citenamefont {Eggenberger}, \citenamefont
  {Ehrenreich}, \citenamefont {Eigm{\"u}ller}, \citenamefont {Erikson},
  \citenamefont {Farmer}, \citenamefont {Feltzing}, \citenamefont
  {Oliveira~Fialho}, \citenamefont {Figueira}, \citenamefont {Forveille},
  \citenamefont {Fridlund}, \citenamefont {Garc{\'\i}a}, \citenamefont
  {Giommi}, \citenamefont {Giuffrida}, \citenamefont {Godolt}, \citenamefont
  {da~Silva}, \citenamefont {Granzer}, \citenamefont {Grenfell}, \citenamefont
  {Grotsch-Noels}, \citenamefont {G{\"u}nther}, \citenamefont {Haswell},
  \citenamefont {Hatzes}, \citenamefont {H{\'e}brard}, \citenamefont {Hekker},
  \citenamefont {Helled}, \citenamefont {Heng}, \citenamefont {Jenkins},
  \citenamefont {Johansen}, \citenamefont {Khodachenko}, \citenamefont
  {Kislyakova}, \citenamefont {Kley}, \citenamefont {Kolb}, \citenamefont
  {Krivova}, \citenamefont {Kupka}, \citenamefont {Lammer}, \citenamefont
  {Lanza}, \citenamefont {Lebreton}, \citenamefont {Magrin}, \citenamefont
  {Marcos-Arenal}, \citenamefont {Marrese}, \citenamefont {Marques},
  \citenamefont {Martins}, \citenamefont {Mathis}, \citenamefont {Mathur},
  \citenamefont {Messina}, \citenamefont {Miglio}, \citenamefont {Montalban},
  \citenamefont {Montalto}, \citenamefont {P.~F. G.~Monteiro}, \citenamefont
  {Moradi}, \citenamefont {Moravveji}, \citenamefont {Mordasini}, \citenamefont
  {Morel}, \citenamefont {Mortier}, \citenamefont {Nascimbeni}, \citenamefont
  {Nelson}, \citenamefont {Nielsen}, \citenamefont {Noack}, \citenamefont
  {Norton}, \citenamefont {Ofir}, \citenamefont {Oshagh}, \citenamefont
  {Ouazzani}, \citenamefont {P{\'a}pics}, \citenamefont {Parro}, \citenamefont
  {Petit}, \citenamefont {Plez}, \citenamefont {Poretti}, \citenamefont
  {Quirrenbach}, \citenamefont {Ragazzoni}, \citenamefont {Raimondo},
  \citenamefont {Rainer}, \citenamefont {Reese}, \citenamefont {Redmer},
  \citenamefont {Reffert}, \citenamefont {Rojas-Ayala}, \citenamefont
  {Roxburgh}, \citenamefont {Salmon}, \citenamefont {Santerne}, \citenamefont
  {Schneider}, \citenamefont {Schou}, \citenamefont {Schuh}, \citenamefont
  {Schunker}, \citenamefont {Silva-Valio}, \citenamefont {Silvotti},
  \citenamefont {Skillen}, \citenamefont {Snellen}, \citenamefont {Sohl},
  \citenamefont {Sousa}, \citenamefont {Sozzetti}, \citenamefont {Stello},
  \citenamefont {Strassmeier}, \citenamefont {{\v S}vanda}, \citenamefont
  {Szab{\'o}}, \citenamefont {Tkachenko}, \citenamefont {Valencia},
  \citenamefont {Van~Grootel}, \citenamefont {Vauclair}, \citenamefont
  {Ventura}, \citenamefont {Wagner}, \citenamefont {Walton}, \citenamefont
  {Weingrill}, \citenamefont {Werner}, \citenamefont {Wheatley},\ and\
  \citenamefont {Zwintz}}]{plato}%
  \BibitemOpen
  \bibfield  {author} {\bibinfo {author} {\bibfnamefont {H.}~\bibnamefont
  {Rauer}}, \bibinfo {author} {\bibfnamefont {C.}~\bibnamefont {Catala}},
  \bibinfo {author} {\bibfnamefont {C.}~\bibnamefont {Aerts}}, \bibinfo
  {author} {\bibfnamefont {T.}~\bibnamefont {Appourchaux}}, \bibinfo {author}
  {\bibfnamefont {W.}~\bibnamefont {Benz}}, \bibinfo {author} {\bibfnamefont
  {A.}~\bibnamefont {Brandeker}}, \bibinfo {author} {\bibfnamefont
  {J.}~\bibnamefont {Christensen-Dalsgaard}}, \bibinfo {author} {\bibfnamefont
  {M.}~\bibnamefont {Deleuil}}, \bibinfo {author} {\bibfnamefont
  {L.}~\bibnamefont {Gizon}}, \bibinfo {author} {\bibfnamefont {M.~J.}\
  \bibnamefont {Goupil}}, \bibinfo {author} {\bibfnamefont {M.}~\bibnamefont
  {G{\"u}del}}, \bibinfo {author} {\bibfnamefont {E.}~\bibnamefont
  {Janot-Pacheco}}, \bibinfo {author} {\bibfnamefont {M.}~\bibnamefont
  {Mas-Hesse}}, \bibinfo {author} {\bibfnamefont {I.}~\bibnamefont {Pagano}},
  \bibinfo {author} {\bibfnamefont {G.}~\bibnamefont {Piotto}}, \bibinfo
  {author} {\bibfnamefont {D.}~\bibnamefont {Pollacco}}, \bibinfo {author}
  {\bibfnamefont {{\.C}.}~\bibnamefont {Santos}}, \bibinfo {author}
  {\bibfnamefont {A.}~\bibnamefont {Smith}}, \bibinfo {author} {\bibfnamefont
  {J.~C.}\ \bibnamefont {Su{\'a}rez}}, \bibinfo {author} {\bibfnamefont
  {R.}~\bibnamefont {Szab{\'o}}}, \bibinfo {author} {\bibfnamefont
  {S.}~\bibnamefont {Udry}}, \bibinfo {author} {\bibfnamefont {V.}~\bibnamefont
  {Adibekyan}}, \bibinfo {author} {\bibfnamefont {Y.}~\bibnamefont {Alibert}},
  \bibinfo {author} {\bibfnamefont {J.~M.}\ \bibnamefont {Almenara}}, \bibinfo
  {author} {\bibfnamefont {P.}~\bibnamefont {Amaro-Seoane}}, \bibinfo {author}
  {\bibfnamefont {M.~A.-v.}\ \bibnamefont {Eiff}}, \bibinfo {author}
  {\bibfnamefont {M.}~\bibnamefont {Asplund}}, \bibinfo {author} {\bibfnamefont
  {E.}~\bibnamefont {Antonello}}, \bibinfo {author} {\bibfnamefont
  {S.}~\bibnamefont {Barnes}}, \bibinfo {author} {\bibfnamefont
  {F.}~\bibnamefont {Baudin}}, \bibinfo {author} {\bibfnamefont
  {K.}~\bibnamefont {Belkacem}}, \bibinfo {author} {\bibfnamefont
  {M.}~\bibnamefont {Bergemann}}, \bibinfo {author} {\bibfnamefont
  {G.}~\bibnamefont {Bihain}}, \bibinfo {author} {\bibfnamefont {A.~C.}\
  \bibnamefont {Birch}}, \bibinfo {author} {\bibfnamefont {X.}~\bibnamefont
  {Bonfils}}, \bibinfo {author} {\bibfnamefont {I.}~\bibnamefont {Boisse}},
  \bibinfo {author} {\bibfnamefont {A.~S.}\ \bibnamefont {Bonomo}}, \bibinfo
  {author} {\bibfnamefont {F.}~\bibnamefont {Borsa}}, \bibinfo {author}
  {\bibfnamefont {I.~M.}\ \bibnamefont {Brand{\~a}o}}, \bibinfo {author}
  {\bibfnamefont {E.}~\bibnamefont {Brocato}}, \bibinfo {author} {\bibfnamefont
  {S.}~\bibnamefont {Brun}}, \bibinfo {author} {\bibfnamefont {M.}~\bibnamefont
  {Burleigh}}, \bibinfo {author} {\bibfnamefont {R.}~\bibnamefont {Burston}},
  \bibinfo {author} {\bibfnamefont {J.}~\bibnamefont {Cabrera}}, \bibinfo
  {author} {\bibfnamefont {S.}~\bibnamefont {Cassisi}}, \bibinfo {author}
  {\bibfnamefont {W.}~\bibnamefont {Chaplin}}, \bibinfo {author} {\bibfnamefont
  {S.}~\bibnamefont {Charpinet}}, \bibinfo {author} {\bibfnamefont
  {C.}~\bibnamefont {Chiappini}}, \bibinfo {author} {\bibfnamefont {R.~P.}\
  \bibnamefont {Church}}, \bibinfo {author} {\bibfnamefont {S.}~\bibnamefont
  {Csizmadia}}, \bibinfo {author} {\bibfnamefont {M.}~\bibnamefont {Cunha}},
  \bibinfo {author} {\bibfnamefont {M.}~\bibnamefont {Damasso}}, \bibinfo
  {author} {\bibfnamefont {M.~B.}\ \bibnamefont {Davies}}, \bibinfo {author}
  {\bibfnamefont {H.~J.}\ \bibnamefont {Deeg}}, \bibinfo {author}
  {\bibfnamefont {R.~F.}\ \bibnamefont {D{\'\i}az}}, \bibinfo {author}
  {\bibfnamefont {S.}~\bibnamefont {Dreizler}}, \bibinfo {author}
  {\bibfnamefont {C.}~\bibnamefont {Dreyer}}, \bibinfo {author} {\bibfnamefont
  {P.}~\bibnamefont {Eggenberger}}, \bibinfo {author} {\bibfnamefont
  {D.}~\bibnamefont {Ehrenreich}}, \bibinfo {author} {\bibfnamefont
  {P.}~\bibnamefont {Eigm{\"u}ller}}, \bibinfo {author} {\bibfnamefont
  {A.}~\bibnamefont {Erikson}}, \bibinfo {author} {\bibfnamefont
  {R.}~\bibnamefont {Farmer}}, \bibinfo {author} {\bibfnamefont
  {S.}~\bibnamefont {Feltzing}}, \bibinfo {author} {\bibfnamefont {F.~d.}\
  \bibnamefont {Oliveira~Fialho}}, \bibinfo {author} {\bibfnamefont
  {P.}~\bibnamefont {Figueira}}, \bibinfo {author} {\bibfnamefont
  {T.}~\bibnamefont {Forveille}}, \bibinfo {author} {\bibfnamefont
  {M.}~\bibnamefont {Fridlund}}, \bibinfo {author} {\bibfnamefont {R.~A.}\
  \bibnamefont {Garc{\'\i}a}}, \bibinfo {author} {\bibfnamefont
  {P.}~\bibnamefont {Giommi}}, \bibinfo {author} {\bibfnamefont
  {G.}~\bibnamefont {Giuffrida}}, \bibinfo {author} {\bibfnamefont
  {M.}~\bibnamefont {Godolt}}, \bibinfo {author} {\bibfnamefont {J.~G.}\
  \bibnamefont {da~Silva}}, \bibinfo {author} {\bibfnamefont {T.}~\bibnamefont
  {Granzer}}, \bibinfo {author} {\bibfnamefont {J.~L.}\ \bibnamefont
  {Grenfell}}, \bibinfo {author} {\bibfnamefont {A.}~\bibnamefont
  {Grotsch-Noels}}, \bibinfo {author} {\bibfnamefont {E.}~\bibnamefont
  {G{\"u}nther}}, \bibinfo {author} {\bibfnamefont {C.~A.}\ \bibnamefont
  {Haswell}}, \bibinfo {author} {\bibfnamefont {A.~P.}\ \bibnamefont {Hatzes}},
  \bibinfo {author} {\bibfnamefont {G.}~\bibnamefont {H{\'e}brard}}, \bibinfo
  {author} {\bibfnamefont {S.}~\bibnamefont {Hekker}}, \bibinfo {author}
  {\bibfnamefont {R.}~\bibnamefont {Helled}}, \bibinfo {author} {\bibfnamefont
  {K.}~\bibnamefont {Heng}}, \bibinfo {author} {\bibfnamefont {J.~M.}\
  \bibnamefont {Jenkins}}, \bibinfo {author} {\bibfnamefont {A.}~\bibnamefont
  {Johansen}}, \bibinfo {author} {\bibfnamefont {M.~L.}\ \bibnamefont
  {Khodachenko}}, \bibinfo {author} {\bibfnamefont {K.~G.}\ \bibnamefont
  {Kislyakova}}, \bibinfo {author} {\bibfnamefont {W.}~\bibnamefont {Kley}},
  \bibinfo {author} {\bibfnamefont {U.}~\bibnamefont {Kolb}}, \bibinfo {author}
  {\bibfnamefont {N.}~\bibnamefont {Krivova}}, \bibinfo {author} {\bibfnamefont
  {F.}~\bibnamefont {Kupka}}, \bibinfo {author} {\bibfnamefont
  {H.}~\bibnamefont {Lammer}}, \bibinfo {author} {\bibfnamefont {A.~F.}\
  \bibnamefont {Lanza}}, \bibinfo {author} {\bibfnamefont {Y.}~\bibnamefont
  {Lebreton}}, \bibinfo {author} {\bibfnamefont {D.}~\bibnamefont {Magrin}},
  \bibinfo {author} {\bibfnamefont {P.}~\bibnamefont {Marcos-Arenal}}, \bibinfo
  {author} {\bibfnamefont {P.~M.}\ \bibnamefont {Marrese}}, \bibinfo {author}
  {\bibfnamefont {J.~P.}\ \bibnamefont {Marques}}, \bibinfo {author}
  {\bibfnamefont {J.}~\bibnamefont {Martins}}, \bibinfo {author} {\bibfnamefont
  {S.}~\bibnamefont {Mathis}}, \bibinfo {author} {\bibfnamefont
  {S.}~\bibnamefont {Mathur}}, \bibinfo {author} {\bibfnamefont
  {S.}~\bibnamefont {Messina}}, \bibinfo {author} {\bibfnamefont
  {A.}~\bibnamefont {Miglio}}, \bibinfo {author} {\bibfnamefont
  {J.}~\bibnamefont {Montalban}}, \bibinfo {author} {\bibfnamefont
  {M.}~\bibnamefont {Montalto}}, \bibinfo {author} {\bibfnamefont {M.~J.}\
  \bibnamefont {P.~F. G.~Monteiro}}, \bibinfo {author} {\bibfnamefont
  {H.}~\bibnamefont {Moradi}}, \bibinfo {author} {\bibfnamefont
  {E.}~\bibnamefont {Moravveji}}, \bibinfo {author} {\bibfnamefont
  {C.}~\bibnamefont {Mordasini}}, \bibinfo {author} {\bibfnamefont
  {T.}~\bibnamefont {Morel}}, \bibinfo {author} {\bibfnamefont
  {A.}~\bibnamefont {Mortier}}, \bibinfo {author} {\bibfnamefont
  {V.}~\bibnamefont {Nascimbeni}}, \bibinfo {author} {\bibfnamefont {R.~P.}\
  \bibnamefont {Nelson}}, \bibinfo {author} {\bibfnamefont {M.~B.}\
  \bibnamefont {Nielsen}}, \bibinfo {author} {\bibfnamefont {L.}~\bibnamefont
  {Noack}}, \bibinfo {author} {\bibfnamefont {A.~J.}\ \bibnamefont {Norton}},
  \bibinfo {author} {\bibfnamefont {A.}~\bibnamefont {Ofir}}, \bibinfo {author}
  {\bibfnamefont {M.}~\bibnamefont {Oshagh}}, \bibinfo {author} {\bibfnamefont
  {R.~M.}\ \bibnamefont {Ouazzani}}, \bibinfo {author} {\bibfnamefont
  {P.}~\bibnamefont {P{\'a}pics}}, \bibinfo {author} {\bibfnamefont {V.~C.}\
  \bibnamefont {Parro}}, \bibinfo {author} {\bibfnamefont {P.}~\bibnamefont
  {Petit}}, \bibinfo {author} {\bibfnamefont {B.}~\bibnamefont {Plez}},
  \bibinfo {author} {\bibfnamefont {E.}~\bibnamefont {Poretti}}, \bibinfo
  {author} {\bibfnamefont {A.}~\bibnamefont {Quirrenbach}}, \bibinfo {author}
  {\bibfnamefont {R.}~\bibnamefont {Ragazzoni}}, \bibinfo {author}
  {\bibfnamefont {G.}~\bibnamefont {Raimondo}}, \bibinfo {author}
  {\bibfnamefont {M.}~\bibnamefont {Rainer}}, \bibinfo {author} {\bibfnamefont
  {D.~R.}\ \bibnamefont {Reese}}, \bibinfo {author} {\bibfnamefont
  {R.}~\bibnamefont {Redmer}}, \bibinfo {author} {\bibfnamefont
  {S.}~\bibnamefont {Reffert}}, \bibinfo {author} {\bibfnamefont
  {B.}~\bibnamefont {Rojas-Ayala}}, \bibinfo {author} {\bibfnamefont {I.~W.}\
  \bibnamefont {Roxburgh}}, \bibinfo {author} {\bibfnamefont {S.}~\bibnamefont
  {Salmon}}, \bibinfo {author} {\bibfnamefont {A.}~\bibnamefont {Santerne}},
  \bibinfo {author} {\bibfnamefont {J.}~\bibnamefont {Schneider}}, \bibinfo
  {author} {\bibfnamefont {J.}~\bibnamefont {Schou}}, \bibinfo {author}
  {\bibfnamefont {S.}~\bibnamefont {Schuh}}, \bibinfo {author} {\bibfnamefont
  {H.}~\bibnamefont {Schunker}}, \bibinfo {author} {\bibfnamefont
  {A.}~\bibnamefont {Silva-Valio}}, \bibinfo {author} {\bibfnamefont
  {R.}~\bibnamefont {Silvotti}}, \bibinfo {author} {\bibfnamefont
  {I.}~\bibnamefont {Skillen}}, \bibinfo {author} {\bibfnamefont
  {I.}~\bibnamefont {Snellen}}, \bibinfo {author} {\bibfnamefont
  {F.}~\bibnamefont {Sohl}}, \bibinfo {author} {\bibfnamefont {S.~G.}\
  \bibnamefont {Sousa}}, \bibinfo {author} {\bibfnamefont {A.}~\bibnamefont
  {Sozzetti}}, \bibinfo {author} {\bibfnamefont {D.}~\bibnamefont {Stello}},
  \bibinfo {author} {\bibfnamefont {K.~G.}\ \bibnamefont {Strassmeier}},
  \bibinfo {author} {\bibfnamefont {M.}~\bibnamefont {{\v S}vanda}}, \bibinfo
  {author} {\bibfnamefont {G.~M.}\ \bibnamefont {Szab{\'o}}}, \bibinfo {author}
  {\bibfnamefont {A.}~\bibnamefont {Tkachenko}}, \bibinfo {author}
  {\bibfnamefont {D.}~\bibnamefont {Valencia}}, \bibinfo {author}
  {\bibfnamefont {V.}~\bibnamefont {Van~Grootel}}, \bibinfo {author}
  {\bibfnamefont {S.~D.}\ \bibnamefont {Vauclair}}, \bibinfo {author}
  {\bibfnamefont {P.}~\bibnamefont {Ventura}}, \bibinfo {author} {\bibfnamefont
  {F.~W.}\ \bibnamefont {Wagner}}, \bibinfo {author} {\bibfnamefont {N.~A.}\
  \bibnamefont {Walton}}, \bibinfo {author} {\bibfnamefont {J.}~\bibnamefont
  {Weingrill}}, \bibinfo {author} {\bibfnamefont {S.~C.}\ \bibnamefont
  {Werner}}, \bibinfo {author} {\bibfnamefont {P.~J.}\ \bibnamefont
  {Wheatley}},\ and\ \bibinfo {author} {\bibfnamefont {K.}~\bibnamefont
  {Zwintz}},\ }\bibfield  {title} {\bibinfo {title} {The plato 2.0 mission},\
  }\href {https://doi.org/10.1007/s10686-014-9383-4} {\bibfield  {journal}
  {\bibinfo  {journal} {Experimental Astronomy}\ }\textbf {\bibinfo {volume}
  {38}},\ \bibinfo {pages} {249} (\bibinfo {year} {2014})}\BibitemShut
  {NoStop}%
\bibitem [{\citenamefont {{Catala}}\ \emph {et~al.}(2010)\citenamefont
  {{Catala}}, \citenamefont {{Arentoft}}, \citenamefont {{Fridlund}},
  \citenamefont {{Lindberg}}, \citenamefont {{Mas-Hesse}}, \citenamefont
  {{Micela}}, \citenamefont {{Pollacco}}, \citenamefont {{Poretti}},
  \citenamefont {{Rauer}}, \citenamefont {{Roxburgh}}, \citenamefont
  {{Stankov}},\ and\ \citenamefont {{Udry}}}]{2010ASPC..430..260C}%
  \BibitemOpen
  \bibfield  {author} {\bibinfo {author} {\bibfnamefont {C.}~\bibnamefont
  {{Catala}}}, \bibinfo {author} {\bibfnamefont {T.}~\bibnamefont
  {{Arentoft}}}, \bibinfo {author} {\bibfnamefont {M.}~\bibnamefont
  {{Fridlund}}}, \bibinfo {author} {\bibfnamefont {R.}~\bibnamefont
  {{Lindberg}}}, \bibinfo {author} {\bibfnamefont {J.~M.}\ \bibnamefont
  {{Mas-Hesse}}}, \bibinfo {author} {\bibfnamefont {G.}~\bibnamefont
  {{Micela}}}, \bibinfo {author} {\bibfnamefont {D.}~\bibnamefont
  {{Pollacco}}}, \bibinfo {author} {\bibfnamefont {E.}~\bibnamefont
  {{Poretti}}}, \bibinfo {author} {\bibfnamefont {H.}~\bibnamefont {{Rauer}}},
  \bibinfo {author} {\bibfnamefont {I.}~\bibnamefont {{Roxburgh}}}, \bibinfo
  {author} {\bibfnamefont {A.}~\bibnamefont {{Stankov}}},\ and\ \bibinfo
  {author} {\bibfnamefont {S.}~\bibnamefont {{Udry}}},\ }\bibfield  {title}
  {\bibinfo {title} {{PLATO : PLAnetary Transits and Oscillations of Stars -
  The Exoplanetary System Explorer}},\ }in\ \href@noop {} {\emph {\bibinfo
  {booktitle} {Pathways Towards Habitable Planets}}},\ \bibinfo {series}
  {Astronomical Society of the Pacific Conference Series}, Vol.\ \bibinfo
  {volume} {430},\ \bibinfo {editor} {edited by\ \bibinfo {editor}
  {\bibfnamefont {V.}~\bibnamefont {{Coud{\'e} du Foresto}}}, \bibinfo {editor}
  {\bibfnamefont {D.~M.}\ \bibnamefont {{Gelino}}},\ and\ \bibinfo {editor}
  {\bibfnamefont {I.}~\bibnamefont {{Ribas}}}}\ (\bibinfo {year} {2010})\ p.\
  \bibinfo {pages} {260}\BibitemShut {NoStop}%
\bibitem [{\citenamefont {Olive}\ \emph {et~al.}(2014)\citenamefont {Olive}
  \emph {et~al.}}]{ParticleDataGroup:2014cgo}%
  \BibitemOpen
  \bibfield  {author} {\bibinfo {author} {\bibfnamefont {K.~A.}\ \bibnamefont
  {Olive}} \emph {et~al.} (\bibinfo {collaboration} {Particle Data Group}),\
  }\bibfield  {title} {\bibinfo {title} {{Review of Particle Physics}},\ }\href
  {https://doi.org/10.1088/1674-1137/38/9/090001} {\bibfield  {journal}
  {\bibinfo  {journal} {Chin. Phys. C}\ }\textbf {\bibinfo {volume} {38}},\
  \bibinfo {pages} {090001} (\bibinfo {year} {2014})}\BibitemShut {NoStop}%
\bibitem [{\citenamefont {Burghoff}\ \emph {et~al.}(2011)\citenamefont
  {Burghoff} \emph {et~al.}}]{Burghoff:2011xk}%
  \BibitemOpen
  \bibfield  {author} {\bibinfo {author} {\bibfnamefont {M.}~\bibnamefont
  {Burghoff}} \emph {et~al.},\ }\bibfield  {title} {\bibinfo {title} {{An
  Improved Search for the Neutron Electric Dipole Moment}},\ }in\ \href@noop {}
  {\emph {\bibinfo {booktitle} {{Meeting of the APS Division of Particles and
  Fields}}}}\ (\bibinfo {year} {2011})\ \Eprint
  {https://arxiv.org/abs/1110.1505} {arXiv:1110.1505 [nucl-ex]} \BibitemShut
  {NoStop}%
\bibitem [{\citenamefont {Ringwald}(2015)}]{Ringwald:2015lqa}%
  \BibitemOpen
  \bibfield  {author} {\bibinfo {author} {\bibfnamefont {A.}~\bibnamefont
  {Ringwald}},\ }\bibfield  {title} {\bibinfo {title} {{The hunt for axions}},\
  }\href {https://doi.org/10.22323/1.244.0021} {\bibfield  {journal} {\bibinfo
  {journal} {PoS}\ }\textbf {\bibinfo {volume} {NEUTEL2015}},\ \bibinfo {pages}
  {021} (\bibinfo {year} {2015})},\ \Eprint {https://arxiv.org/abs/1506.04259}
  {arXiv:1506.04259 [hep-ph]} \BibitemShut {NoStop}%
\bibitem [{\citenamefont {Raffelt}(1986)}]{Raffelt:1985nk}%
  \BibitemOpen
  \bibfield  {author} {\bibinfo {author} {\bibfnamefont {G.~G.}\ \bibnamefont
  {Raffelt}},\ }\bibfield  {title} {\bibinfo {title} {{ASTROPHYSICAL AXION
  BOUNDS DIMINISHED BY SCREENING EFFECTS}},\ }\href
  {https://doi.org/10.1103/PhysRevD.33.897} {\bibfield  {journal} {\bibinfo
  {journal} {Phys. Rev. D}\ }\textbf {\bibinfo {volume} {33}},\ \bibinfo
  {pages} {897} (\bibinfo {year} {1986})}\BibitemShut {NoStop}%
\bibitem [{\citenamefont {Kim}(1979)}]{Kim:1979if}%
  \BibitemOpen
  \bibfield  {author} {\bibinfo {author} {\bibfnamefont {J.~E.}\ \bibnamefont
  {Kim}},\ }\bibfield  {title} {\bibinfo {title} {{Weak Interaction Singlet and
  Strong CP Invariance}},\ }\href {https://doi.org/10.1103/PhysRevLett.43.103}
  {\bibfield  {journal} {\bibinfo  {journal} {Phys. Rev. Lett.}\ }\textbf
  {\bibinfo {volume} {43}},\ \bibinfo {pages} {103} (\bibinfo {year}
  {1979})}\BibitemShut {NoStop}%
\bibitem [{\citenamefont {Shifman}\ \emph {et~al.}(1980)\citenamefont
  {Shifman}, \citenamefont {Vainshtein},\ and\ \citenamefont
  {Zakharov}}]{Shifman:1979if}%
  \BibitemOpen
  \bibfield  {author} {\bibinfo {author} {\bibfnamefont {M.~A.}\ \bibnamefont
  {Shifman}}, \bibinfo {author} {\bibfnamefont {A.~I.}\ \bibnamefont
  {Vainshtein}},\ and\ \bibinfo {author} {\bibfnamefont {V.~I.}\ \bibnamefont
  {Zakharov}},\ }\bibfield  {title} {\bibinfo {title} {{Can Confinement Ensure
  Natural CP Invariance of Strong Interactions?}},\ }\href
  {https://doi.org/10.1016/0550-3213(80)90209-6} {\bibfield  {journal}
  {\bibinfo  {journal} {Nucl. Phys. B}\ }\textbf {\bibinfo {volume} {166}},\
  \bibinfo {pages} {493} (\bibinfo {year} {1980})}\BibitemShut {NoStop}%
\bibitem [{\citenamefont {Raffelt}(1999)}]{Raffelt:1998fy}%
  \BibitemOpen
  \bibfield  {author} {\bibinfo {author} {\bibfnamefont {G.}~\bibnamefont
  {Raffelt}},\ }\bibfield  {title} {\bibinfo {title} {{Stellar evolution limits
  on axion properties}},\ }\href
  {https://doi.org/10.1016/S0920-5632(98)00501-5} {\bibfield  {journal}
  {\bibinfo  {journal} {Nucl. Phys. B Proc. Suppl.}\ }\textbf {\bibinfo
  {volume} {72}},\ \bibinfo {pages} {43} (\bibinfo {year} {1999})},\ \Eprint
  {https://arxiv.org/abs/hep-ph/9805400} {arXiv:hep-ph/9805400} \BibitemShut
  {NoStop}%
\bibitem [{\citenamefont {Li}(2020)}]{Li:2020naa}%
  \BibitemOpen
  \bibfield  {author} {\bibinfo {author} {\bibfnamefont {T.}~\bibnamefont
  {Li}},\ }\bibfield  {title} {\bibinfo {title} {{The KSVZ Axion and
  Pseudo-Nambu-Goldstone Boson Models for the XENON1T Excess}},\ }\href@noop {}
  {\  (\bibinfo {year} {2020})},\ \Eprint {https://arxiv.org/abs/2007.00874}
  {arXiv:2007.00874 [hep-ph]} \BibitemShut {NoStop}%
\bibitem [{\citenamefont {H{\"u}ckel}(2015)}]{Huckel1924}%
  \BibitemOpen
  \bibfield  {author} {\bibinfo {author} {\bibfnamefont {E.}~\bibnamefont
  {H{\"u}ckel}},\ }\bibfield  {title} {\bibinfo {title} {Zur theorie der
  elektrolyte},\ }\bibfield  {journal} {\bibinfo  {journal} {Phys. Z.}\
  }\textbf {\bibinfo {volume} {24}},\ \href
  {https://doi.org/10.1007/BFb0111753} {10.1007/BFb0111753} (\bibinfo {year}
  {2015})\BibitemShut {NoStop}%
\bibitem [{\citenamefont {Raffelt}(1990)}]{Raffelt:1990yz}%
  \BibitemOpen
  \bibfield  {author} {\bibinfo {author} {\bibfnamefont {G.~G.}\ \bibnamefont
  {Raffelt}},\ }\bibfield  {title} {\bibinfo {title} {{Astrophysical methods to
  constrain axions and other novel particle phenomena}},\ }\href
  {https://doi.org/10.1016/0370-1573(90)90054-6} {\bibfield  {journal}
  {\bibinfo  {journal} {Phys. Rept.}\ }\textbf {\bibinfo {volume} {198}},\
  \bibinfo {pages} {1} (\bibinfo {year} {1990})}\BibitemShut {NoStop}%
\bibitem [{\citenamefont {Choplin}\ \emph {et~al.}(2017)\citenamefont
  {Choplin}, \citenamefont {Coc}, \citenamefont {Meynet}, \citenamefont
  {Olive}, \citenamefont {Uzan},\ and\ \citenamefont
  {Vangioni}}]{Choplin:2017auq}%
  \BibitemOpen
  \bibfield  {author} {\bibinfo {author} {\bibfnamefont {A.}~\bibnamefont
  {Choplin}}, \bibinfo {author} {\bibfnamefont {A.}~\bibnamefont {Coc}},
  \bibinfo {author} {\bibfnamefont {G.}~\bibnamefont {Meynet}}, \bibinfo
  {author} {\bibfnamefont {K.~A.}\ \bibnamefont {Olive}}, \bibinfo {author}
  {\bibfnamefont {J.-P.}\ \bibnamefont {Uzan}},\ and\ \bibinfo {author}
  {\bibfnamefont {E.}~\bibnamefont {Vangioni}},\ }\bibfield  {title} {\bibinfo
  {title} {{Effects of axions on Population III stars}},\ }\href
  {https://doi.org/10.1051/0004-6361/201731040} {\bibfield  {journal} {\bibinfo
   {journal} {Astron. Astrophys.}\ }\textbf {\bibinfo {volume} {605}},\
  \bibinfo {pages} {A106} (\bibinfo {year} {2017})},\ \Eprint
  {https://arxiv.org/abs/1707.01244} {arXiv:1707.01244 [astro-ph.SR]}
  \BibitemShut {NoStop}%
\bibitem [{\citenamefont {Friedland}\ \emph
  {et~al.}(2013{\natexlab{a}})\citenamefont {Friedland}, \citenamefont
  {Giannotti},\ and\ \citenamefont {Wise}}]{Friedland:2013fse}%
  \BibitemOpen
  \bibfield  {author} {\bibinfo {author} {\bibfnamefont {A.}~\bibnamefont
  {Friedland}}, \bibinfo {author} {\bibfnamefont {M.}~\bibnamefont
  {Giannotti}},\ and\ \bibinfo {author} {\bibfnamefont {M.}~\bibnamefont
  {Wise}},\ }\bibfield  {title} {\bibinfo {title} {{A new Constraint on the
  Axion-Photon Coupling}},\ }in\ \href
  {https://doi.org/10.3204/DESY-PROC-2013-04/giannotti_maurizio} {\emph
  {\bibinfo {booktitle} {{9th Patras Workshop on Axions, WIMPs and WISPs}}}}\
  (\bibinfo {year} {2013})\ pp.\ \bibinfo {pages} {73--76}\BibitemShut
  {NoStop}%
\bibitem [{\citenamefont {Agrawal}\ \emph {et~al.}(2020)\citenamefont
  {Agrawal}, \citenamefont {Hurley}, \citenamefont {Stevenson}, \citenamefont
  {Sz\'ecsi},\ and\ \citenamefont {Flynn}}]{Agrawal:2020znh}%
  \BibitemOpen
  \bibfield  {author} {\bibinfo {author} {\bibfnamefont {P.}~\bibnamefont
  {Agrawal}}, \bibinfo {author} {\bibfnamefont {J.}~\bibnamefont {Hurley}},
  \bibinfo {author} {\bibfnamefont {S.}~\bibnamefont {Stevenson}}, \bibinfo
  {author} {\bibfnamefont {D.}~\bibnamefont {Sz\'ecsi}},\ and\ \bibinfo
  {author} {\bibfnamefont {C.}~\bibnamefont {Flynn}},\ }\bibfield  {title}
  {\bibinfo {title} {{The fates of massive stars: exploring uncertainties in
  stellar evolution with METISSE}},\ }\href
  {https://doi.org/10.1093/mnras/staa2264} {\bibfield  {journal} {\bibinfo
  {journal} {Mon. Not. Roy. Astron. Soc.}\ }\textbf {\bibinfo {volume} {497}},\
  \bibinfo {pages} {4549} (\bibinfo {year} {2020})},\ \Eprint
  {https://arxiv.org/abs/2005.13177} {arXiv:2005.13177 [astro-ph.SR]}
  \BibitemShut {NoStop}%
\bibitem [{\citenamefont {Rato}\ \emph {et~al.}(2021)\citenamefont {Rato},
  \citenamefont {Lopes},\ and\ \citenamefont {Lopes}}]{Rato:2021tfc}%
  \BibitemOpen
  \bibfield  {author} {\bibinfo {author} {\bibfnamefont {J.~a.}\ \bibnamefont
  {Rato}}, \bibinfo {author} {\bibfnamefont {J.}~\bibnamefont {Lopes}},\ and\
  \bibinfo {author} {\bibfnamefont {I.}~\bibnamefont {Lopes}},\ }\bibfield
  {title} {\bibinfo {title} {{On asymmetric dark matter constraints from the
  asteroseismology of a subgiant star}},\ }\href
  {https://doi.org/10.1093/mnras/stab2372} {\bibfield  {journal} {\bibinfo
  {journal} {Mon. Not. Roy. Astron. Soc.}\ }\textbf {\bibinfo {volume} {507}},\
  \bibinfo {pages} {3434} (\bibinfo {year} {2021})},\ \Eprint
  {https://arxiv.org/abs/2109.12671} {arXiv:2109.12671 [astro-ph.SR]}
  \BibitemShut {NoStop}%
\bibitem [{\citenamefont {{Aerts}}\ \emph {et~al.}(2010)\citenamefont
  {{Aerts}}, \citenamefont {{Christensen-Dalsgaard}},\ and\ \citenamefont
  {{Kurtz}}}]{dalsgaard}%
  \BibitemOpen
  \bibfield  {author} {\bibinfo {author} {\bibfnamefont {C.}~\bibnamefont
  {{Aerts}}}, \bibinfo {author} {\bibfnamefont {J.}~\bibnamefont
  {{Christensen-Dalsgaard}}},\ and\ \bibinfo {author} {\bibfnamefont {D.~W.}\
  \bibnamefont {{Kurtz}}},\ }\href {https://doi.org/10.1007/978-1-4020-5803-5}
  {\emph {\bibinfo {title} {{Asteroseismology}}}}\ (\bibinfo  {publisher}
  {Springer},\ \bibinfo {year} {2010})\BibitemShut {NoStop}%
\bibitem [{\citenamefont {{Tassoul}}(1980)}]{1980ApJS...43..469T}%
  \BibitemOpen
  \bibfield  {author} {\bibinfo {author} {\bibfnamefont {M.}~\bibnamefont
  {{Tassoul}}},\ }\bibfield  {title} {\bibinfo {title} {{Asymptotic
  approximations for stellar nonradial pulsations.}},\ }\href
  {https://doi.org/10.1086/190678} {\bibfield  {journal} {\bibinfo  {journal}
  {{ApJ}}\ }\textbf {\bibinfo {volume} {43}},\ \bibinfo {pages} {469} (\bibinfo
  {year} {1980})}\BibitemShut {NoStop}%
\bibitem [{\citenamefont {{Creevey, O. L.}}\ \emph {et~al.}(2017)\citenamefont
  {{Creevey, O. L.}}, \citenamefont {{Metcalfe, T. S.}}, \citenamefont
  {{Schultheis, M.}}, \citenamefont {{Salabert, D.}}, \citenamefont {{Bazot,
  M.}}, \citenamefont {{Th\'evenin, F.}}, \citenamefont {{Mathur, S.}},
  \citenamefont {{Xu, H.}},\ and\ \citenamefont {{Garc\'{\i}a, R.
  A.}}}]{refId1}%
  \BibitemOpen
  \bibfield  {author} {\bibinfo {author} {\bibnamefont {{Creevey, O. L.}}},
  \bibinfo {author} {\bibnamefont {{Metcalfe, T. S.}}}, \bibinfo {author}
  {\bibnamefont {{Schultheis, M.}}}, \bibinfo {author} {\bibnamefont
  {{Salabert, D.}}}, \bibinfo {author} {\bibnamefont {{Bazot, M.}}}, \bibinfo
  {author} {\bibnamefont {{Th\'evenin, F.}}}, \bibinfo {author} {\bibnamefont
  {{Mathur, S.}}}, \bibinfo {author} {\bibnamefont {{Xu, H.}}},\ and\ \bibinfo
  {author} {\bibnamefont {{Garc\'{\i}a, R. A.}}},\ }\bibfield  {title}
  {\bibinfo {title} {Characterizing solar-type stars from full-length kepler
  data sets using the asteroseismic modeling portal},\ }\href
  {https://doi.org/10.1051/0004-6361/201629496} {\bibfield  {journal} {\bibinfo
   {journal} {A\&A}\ }\textbf {\bibinfo {volume} {601}},\ \bibinfo {pages}
  {A67} (\bibinfo {year} {2017})}\BibitemShut {NoStop}%
\bibitem [{\citenamefont {{Roxburgh, I. W.}}\ and\ \citenamefont {{Vorontsov,
  S. V.}}(2003)}]{refId0}%
  \BibitemOpen
  \bibfield  {author} {\bibinfo {author} {\bibnamefont {{Roxburgh, I. W.}}}\
  and\ \bibinfo {author} {\bibnamefont {{Vorontsov, S. V.}}},\ }\bibfield
  {title} {\bibinfo {title} {The ratio of small to large separations of
  acoustic oscillations as a diagnostic of the interior of solar-like stars},\
  }\href {https://doi.org/10.1051/0004-6361:20031318} {\bibfield  {journal}
  {\bibinfo  {journal} {A\&A}\ }\textbf {\bibinfo {volume} {411}},\ \bibinfo
  {pages} {215} (\bibinfo {year} {2003})}\BibitemShut {NoStop}%
\bibitem [{\citenamefont {Appourchaux}\ \emph {et~al.}(2012)\citenamefont
  {Appourchaux} \emph {et~al.}}]{Appourchaux:2012zm}%
  \BibitemOpen
  \bibfield  {author} {\bibinfo {author} {\bibfnamefont {T.}~\bibnamefont
  {Appourchaux}} \emph {et~al.},\ }\bibfield  {title} {\bibinfo {title}
  {{Oscillation mode frequencies of 61 main-sequence and subgiant stars
  observed by Kepler}},\ }\href {https://doi.org/10.1051/0004-6361/201218948}
  {\bibfield  {journal} {\bibinfo  {journal} {Astron. Astrophys.}\ }\textbf
  {\bibinfo {volume} {543}},\ \bibinfo {pages} {A54} (\bibinfo {year}
  {2012})},\ \Eprint {https://arxiv.org/abs/1204.3147} {arXiv:1204.3147
  [astro-ph.SR]} \BibitemShut {NoStop}%
\bibitem [{\citenamefont {Molenda-Zakowicz}\ \emph {et~al.}(2013)\citenamefont
  {Molenda-Zakowicz} \emph {et~al.}}]{Molenda-Zakowicz:2013waa}%
  \BibitemOpen
  \bibfield  {author} {\bibinfo {author} {\bibfnamefont {J.}~\bibnamefont
  {Molenda-Zakowicz}} \emph {et~al.},\ }\bibfield  {title} {\bibinfo {title}
  {{Atmospheric Parameters of 169 F, G, K and M-type Stars in the Kepler
  Field}},\ }\href {https://doi.org/10.1093/mnras/stt1095} {\bibfield
  {journal} {\bibinfo  {journal} {Mon. Not. Roy. Astron. Soc.}\ }\textbf
  {\bibinfo {volume} {434}},\ \bibinfo {pages} {1422} (\bibinfo {year}
  {2013})},\ \Eprint {https://arxiv.org/abs/1306.6011} {arXiv:1306.6011
  [astro-ph.SR]} \BibitemShut {NoStop}%
\bibitem [{\citenamefont {Mathur}\ \emph {et~al.}(2012)\citenamefont {Mathur}
  \emph {et~al.}}]{Mathur:2012sk}%
  \BibitemOpen
  \bibfield  {author} {\bibinfo {author} {\bibfnamefont {S.}~\bibnamefont
  {Mathur}} \emph {et~al.},\ }\bibfield  {title} {\bibinfo {title} {{A uniform
  asteroseismic analysis of 22 solar-type stars observed by Kepler}},\ }\href
  {https://doi.org/10.1088/0004-637X/749/2/152} {\bibfield  {journal} {\bibinfo
   {journal} {Astrophys. J.}\ }\textbf {\bibinfo {volume} {749}},\ \bibinfo
  {pages} {152} (\bibinfo {year} {2012})},\ \Eprint
  {https://arxiv.org/abs/1202.2844} {arXiv:1202.2844 [astro-ph.SR]}
  \BibitemShut {NoStop}%
\bibitem [{\citenamefont {{Paxton}}\ \emph {et~al.}(2011)\citenamefont
  {{Paxton}}, \citenamefont {{Bildsten}}, \citenamefont {{Dotter}},
  \citenamefont {{Herwig}}, \citenamefont {{Lesaffre}},\ and\ \citenamefont
  {{Timmes}}}]{Paxton2011}%
  \BibitemOpen
  \bibfield  {author} {\bibinfo {author} {\bibfnamefont {B.}~\bibnamefont
  {{Paxton}}}, \bibinfo {author} {\bibfnamefont {L.}~\bibnamefont
  {{Bildsten}}}, \bibinfo {author} {\bibfnamefont {A.}~\bibnamefont
  {{Dotter}}}, \bibinfo {author} {\bibfnamefont {F.}~\bibnamefont {{Herwig}}},
  \bibinfo {author} {\bibfnamefont {P.}~\bibnamefont {{Lesaffre}}},\ and\
  \bibinfo {author} {\bibfnamefont {F.}~\bibnamefont {{Timmes}}},\ }\bibfield
  {title} {\bibinfo {title} {{Modules for Experiments in Stellar Astrophysics
  (MESA)}},\ }\href {https://doi.org/10.1088/0067-0049/192/1/3} {\bibfield
  {journal} {\bibinfo  {journal} {{ApJ}}\ }\textbf {\bibinfo {volume} {192}},\
  \bibinfo {eid} {3} (\bibinfo {year} {2011})},\ \Eprint
  {https://arxiv.org/abs/1009.1622} {arXiv:1009.1622 [astro-ph.SR]}
  \BibitemShut {NoStop}%
\bibitem [{\citenamefont {{Paxton}}\ \emph {et~al.}(2013)\citenamefont
  {{Paxton}}, \citenamefont {{Cantiello}}, \citenamefont {{Arras}},
  \citenamefont {{Bildsten}}, \citenamefont {{Brown}}, \citenamefont
  {{Dotter}}, \citenamefont {{Mankovich}}, \citenamefont {{Montgomery}},
  \citenamefont {{Stello}}, \citenamefont {{Timmes}},\ and\ \citenamefont
  {{Townsend}}}]{Paxton2013}%
  \BibitemOpen
  \bibfield  {author} {\bibinfo {author} {\bibfnamefont {B.}~\bibnamefont
  {{Paxton}}}, \bibinfo {author} {\bibfnamefont {M.}~\bibnamefont
  {{Cantiello}}}, \bibinfo {author} {\bibfnamefont {P.}~\bibnamefont
  {{Arras}}}, \bibinfo {author} {\bibfnamefont {L.}~\bibnamefont {{Bildsten}}},
  \bibinfo {author} {\bibfnamefont {E.~F.}\ \bibnamefont {{Brown}}}, \bibinfo
  {author} {\bibfnamefont {A.}~\bibnamefont {{Dotter}}}, \bibinfo {author}
  {\bibfnamefont {C.}~\bibnamefont {{Mankovich}}}, \bibinfo {author}
  {\bibfnamefont {M.~H.}\ \bibnamefont {{Montgomery}}}, \bibinfo {author}
  {\bibfnamefont {D.}~\bibnamefont {{Stello}}}, \bibinfo {author}
  {\bibfnamefont {F.~X.}\ \bibnamefont {{Timmes}}},\ and\ \bibinfo {author}
  {\bibfnamefont {R.}~\bibnamefont {{Townsend}}},\ }\bibfield  {title}
  {\bibinfo {title} {{Modules for Experiments in Stellar Astrophysics (MESA):
  Planets, Oscillations, Rotation, and Massive Stars}},\ }\href
  {https://doi.org/10.1088/0067-0049/208/1/4} {\bibfield  {journal} {\bibinfo
  {journal} {{ApJ}}\ }\textbf {\bibinfo {volume} {208}},\ \bibinfo {eid} {4}
  (\bibinfo {year} {2013})},\ \Eprint {https://arxiv.org/abs/1301.0319}
  {arXiv:1301.0319 [astro-ph.SR]} \BibitemShut {NoStop}%
\bibitem [{\citenamefont {{Paxton}}\ \emph {et~al.}(2015)\citenamefont
  {{Paxton}}, \citenamefont {{Marchant}}, \citenamefont {{Schwab}},
  \citenamefont {{Bauer}}, \citenamefont {{Bildsten}}, \citenamefont
  {{Cantiello}}, \citenamefont {{Dessart}}, \citenamefont {{Farmer}},
  \citenamefont {{Hu}}, \citenamefont {{Langer}}, \citenamefont {{Townsend}},
  \citenamefont {{Townsley}},\ and\ \citenamefont {{Timmes}}}]{Paxton2015}%
  \BibitemOpen
  \bibfield  {author} {\bibinfo {author} {\bibfnamefont {B.}~\bibnamefont
  {{Paxton}}}, \bibinfo {author} {\bibfnamefont {P.}~\bibnamefont
  {{Marchant}}}, \bibinfo {author} {\bibfnamefont {J.}~\bibnamefont
  {{Schwab}}}, \bibinfo {author} {\bibfnamefont {E.~B.}\ \bibnamefont
  {{Bauer}}}, \bibinfo {author} {\bibfnamefont {L.}~\bibnamefont {{Bildsten}}},
  \bibinfo {author} {\bibfnamefont {M.}~\bibnamefont {{Cantiello}}}, \bibinfo
  {author} {\bibfnamefont {L.}~\bibnamefont {{Dessart}}}, \bibinfo {author}
  {\bibfnamefont {R.}~\bibnamefont {{Farmer}}}, \bibinfo {author}
  {\bibfnamefont {H.}~\bibnamefont {{Hu}}}, \bibinfo {author} {\bibfnamefont
  {N.}~\bibnamefont {{Langer}}}, \bibinfo {author} {\bibfnamefont {R.~H.~D.}\
  \bibnamefont {{Townsend}}}, \bibinfo {author} {\bibfnamefont {D.~M.}\
  \bibnamefont {{Townsley}}},\ and\ \bibinfo {author} {\bibfnamefont {F.~X.}\
  \bibnamefont {{Timmes}}},\ }\bibfield  {title} {\bibinfo {title} {{Modules
  for Experiments in Stellar Astrophysics (MESA): Binaries, Pulsations, and
  Explosions}},\ }\href {https://doi.org/10.1088/0067-0049/220/1/15} {\bibfield
   {journal} {\bibinfo  {journal} {{ApJ}}\ }\textbf {\bibinfo {volume} {220}},\
  \bibinfo {eid} {15} (\bibinfo {year} {2015})},\ \Eprint
  {https://arxiv.org/abs/1506.03146} {arXiv:1506.03146 [astro-ph.SR]}
  \BibitemShut {NoStop}%
\bibitem [{\citenamefont {{Paxton}}\ \emph {et~al.}(2018)\citenamefont
  {{Paxton}}, \citenamefont {{Schwab}}, \citenamefont {{Bauer}}, \citenamefont
  {{Bildsten}}, \citenamefont {{Blinnikov}}, \citenamefont {{Duffell}},
  \citenamefont {{Farmer}}, \citenamefont {{Goldberg}}, \citenamefont
  {{Marchant}}, \citenamefont {{Sorokina}}, \citenamefont {{Thoul}},
  \citenamefont {{Townsend}},\ and\ \citenamefont {{Timmes}}}]{Paxton2018}%
  \BibitemOpen
  \bibfield  {author} {\bibinfo {author} {\bibfnamefont {B.}~\bibnamefont
  {{Paxton}}}, \bibinfo {author} {\bibfnamefont {J.}~\bibnamefont {{Schwab}}},
  \bibinfo {author} {\bibfnamefont {E.~B.}\ \bibnamefont {{Bauer}}}, \bibinfo
  {author} {\bibfnamefont {L.}~\bibnamefont {{Bildsten}}}, \bibinfo {author}
  {\bibfnamefont {S.}~\bibnamefont {{Blinnikov}}}, \bibinfo {author}
  {\bibfnamefont {P.}~\bibnamefont {{Duffell}}}, \bibinfo {author}
  {\bibfnamefont {R.}~\bibnamefont {{Farmer}}}, \bibinfo {author}
  {\bibfnamefont {J.~A.}\ \bibnamefont {{Goldberg}}}, \bibinfo {author}
  {\bibfnamefont {P.}~\bibnamefont {{Marchant}}}, \bibinfo {author}
  {\bibfnamefont {E.}~\bibnamefont {{Sorokina}}}, \bibinfo {author}
  {\bibfnamefont {A.}~\bibnamefont {{Thoul}}}, \bibinfo {author} {\bibfnamefont
  {R.~H.~D.}\ \bibnamefont {{Townsend}}},\ and\ \bibinfo {author}
  {\bibfnamefont {F.~X.}\ \bibnamefont {{Timmes}}},\ }\bibfield  {title}
  {\bibinfo {title} {{Modules for Experiments in Stellar Astrophysics (MESA):
  Convective Boundaries, Element Diffusion, and Massive Star Explosions}},\
  }\href {https://doi.org/10.3847/1538-4365/aaa5a8} {\bibfield  {journal}
  {\bibinfo  {journal} {{ApJ}}\ }\textbf {\bibinfo {volume} {234}},\ \bibinfo
  {eid} {34} (\bibinfo {year} {2018})},\ \Eprint
  {https://arxiv.org/abs/1710.08424} {arXiv:1710.08424 [astro-ph.SR]}
  \BibitemShut {NoStop}%
\bibitem [{\citenamefont {{Paxton}}\ \emph {et~al.}(2019)\citenamefont
  {{Paxton}}, \citenamefont {{Smolec}}, \citenamefont {{Schwab}}, \citenamefont
  {{Gautschy}}, \citenamefont {{Bildsten}}, \citenamefont {{Cantiello}},
  \citenamefont {{Dotter}}, \citenamefont {{Farmer}}, \citenamefont
  {{Goldberg}}, \citenamefont {{Jermyn}}, \citenamefont {{Kanbur}},
  \citenamefont {{Marchant}}, \citenamefont {{Thoul}}, \citenamefont
  {{Townsend}}, \citenamefont {{Wolf}}, \citenamefont {{Zhang}},\ and\
  \citenamefont {{Timmes}}}]{Paxton2019}%
  \BibitemOpen
  \bibfield  {author} {\bibinfo {author} {\bibfnamefont {B.}~\bibnamefont
  {{Paxton}}}, \bibinfo {author} {\bibfnamefont {R.}~\bibnamefont {{Smolec}}},
  \bibinfo {author} {\bibfnamefont {J.}~\bibnamefont {{Schwab}}}, \bibinfo
  {author} {\bibfnamefont {A.}~\bibnamefont {{Gautschy}}}, \bibinfo {author}
  {\bibfnamefont {L.}~\bibnamefont {{Bildsten}}}, \bibinfo {author}
  {\bibfnamefont {M.}~\bibnamefont {{Cantiello}}}, \bibinfo {author}
  {\bibfnamefont {A.}~\bibnamefont {{Dotter}}}, \bibinfo {author}
  {\bibfnamefont {R.}~\bibnamefont {{Farmer}}}, \bibinfo {author}
  {\bibfnamefont {J.~A.}\ \bibnamefont {{Goldberg}}}, \bibinfo {author}
  {\bibfnamefont {A.~S.}\ \bibnamefont {{Jermyn}}}, \bibinfo {author}
  {\bibfnamefont {S.~M.}\ \bibnamefont {{Kanbur}}}, \bibinfo {author}
  {\bibfnamefont {P.}~\bibnamefont {{Marchant}}}, \bibinfo {author}
  {\bibfnamefont {A.}~\bibnamefont {{Thoul}}}, \bibinfo {author} {\bibfnamefont
  {R.~H.~D.}\ \bibnamefont {{Townsend}}}, \bibinfo {author} {\bibfnamefont
  {W.~M.}\ \bibnamefont {{Wolf}}}, \bibinfo {author} {\bibfnamefont
  {M.}~\bibnamefont {{Zhang}}},\ and\ \bibinfo {author} {\bibfnamefont {F.~X.}\
  \bibnamefont {{Timmes}}},\ }\bibfield  {title} {\bibinfo {title} {{Modules
  for Experiments in Stellar Astrophysics (MESA): Pulsating Variable Stars,
  Rotation, Convective Boundaries, and Energy Conservation}},\ }\href
  {https://doi.org/10.3847/1538-4365/ab2241} {\bibfield  {journal} {\bibinfo
  {journal} {{ApJ}}\ }\textbf {\bibinfo {volume} {243}},\ \bibinfo {eid} {10}
  (\bibinfo {year} {2019})},\ \Eprint {https://arxiv.org/abs/1903.01426}
  {arXiv:1903.01426 [astro-ph.SR]} \BibitemShut {NoStop}%
\bibitem [{\citenamefont {{Rogers}}\ and\ \citenamefont
  {{Nayfonov}}(2002)}]{Rogers2002}%
  \BibitemOpen
  \bibfield  {author} {\bibinfo {author} {\bibfnamefont {F.~J.}\ \bibnamefont
  {{Rogers}}}\ and\ \bibinfo {author} {\bibfnamefont {A.}~\bibnamefont
  {{Nayfonov}}},\ }\bibfield  {title} {\bibinfo {title} {{Updated and Expanded
  OPAL Equation-of-State Tables: Implications for Helioseismology}},\ }\href
  {https://doi.org/10.1086/341894} {\bibfield  {journal} {\bibinfo  {journal}
  {\apj}\ }\textbf {\bibinfo {volume} {576}},\ \bibinfo {pages} {1064}
  (\bibinfo {year} {2002})}\BibitemShut {NoStop}%
\bibitem [{\citenamefont {{Saumon}}\ \emph {et~al.}(1995)\citenamefont
  {{Saumon}}, \citenamefont {{Chabrier}},\ and\ \citenamefont {{van
  Horn}}}]{Saumon1995}%
  \BibitemOpen
  \bibfield  {author} {\bibinfo {author} {\bibfnamefont {D.}~\bibnamefont
  {{Saumon}}}, \bibinfo {author} {\bibfnamefont {G.}~\bibnamefont
  {{Chabrier}}},\ and\ \bibinfo {author} {\bibfnamefont {H.~M.}\ \bibnamefont
  {{van Horn}}},\ }\bibfield  {title} {\bibinfo {title} {{An Equation of State
  for Low-Mass Stars and Giant Planets}},\ }\href
  {https://doi.org/10.1086/192204} {\bibfield  {journal} {\bibinfo  {journal}
  {{ApJ}}\ }\textbf {\bibinfo {volume} {99}},\ \bibinfo {pages} {713} (\bibinfo
  {year} {1995})}\BibitemShut {NoStop}%
\bibitem [{\citenamefont {{Irwin}}(2004)}]{Irwin2004}%
  \BibitemOpen
  \bibfield  {author} {\bibinfo {author} {\bibfnamefont {A.~W.}\ \bibnamefont
  {{Irwin}}},\ }\href {http://freeeos.sourceforge.net/} {\bibinfo {title} {The
  freeeos code for calculating the equation of state for stellar interiors}}
  (\bibinfo {year} {2004})\BibitemShut {NoStop}%
\bibitem [{\citenamefont {{Timmes}}\ and\ \citenamefont
  {{Swesty}}(2000)}]{Timmes2000}%
  \BibitemOpen
  \bibfield  {author} {\bibinfo {author} {\bibfnamefont {F.~X.}\ \bibnamefont
  {{Timmes}}}\ and\ \bibinfo {author} {\bibfnamefont {F.~D.}\ \bibnamefont
  {{Swesty}}},\ }\bibfield  {title} {\bibinfo {title} {{The Accuracy,
  Consistency, and Speed of an Electron-Positron Equation of State Based on
  Table Interpolation of the Helmholtz Free Energy}},\ }\href
  {https://doi.org/10.1086/313304} {\bibfield  {journal} {\bibinfo  {journal}
  {{ApJ}}\ }\textbf {\bibinfo {volume} {126}},\ \bibinfo {pages} {501}
  (\bibinfo {year} {2000})}\BibitemShut {NoStop}%
\bibitem [{\citenamefont {{Potekhin}}\ and\ \citenamefont
  {{Chabrier}}(2010)}]{Potekhin2010}%
  \BibitemOpen
  \bibfield  {author} {\bibinfo {author} {\bibfnamefont {A.~Y.}\ \bibnamefont
  {{Potekhin}}}\ and\ \bibinfo {author} {\bibfnamefont {G.}~\bibnamefont
  {{Chabrier}}},\ }\bibfield  {title} {\bibinfo {title} {{Thermodynamic
  Functions of Dense Plasmas: Analytic Approximations for Astrophysical
  Applications}},\ }\href {https://doi.org/10.1002/ctpp.201010017} {\bibfield
  {journal} {\bibinfo  {journal} {Contributions to Plasma Physics}\ }\textbf
  {\bibinfo {volume} {50}},\ \bibinfo {pages} {82} (\bibinfo {year} {2010})},\
  \Eprint {https://arxiv.org/abs/1001.0690} {arXiv:1001.0690
  [physics.plasm-ph]} \BibitemShut {NoStop}%
\bibitem [{\citenamefont {{Jermyn}}\ \emph {et~al.}(2021)\citenamefont
  {{Jermyn}}, \citenamefont {{Schwab}}, \citenamefont {{Bauer}}, \citenamefont
  {{Timmes}},\ and\ \citenamefont {{Potekhin}}}]{Jermyn2021}%
  \BibitemOpen
  \bibfield  {author} {\bibinfo {author} {\bibfnamefont {A.~S.}\ \bibnamefont
  {{Jermyn}}}, \bibinfo {author} {\bibfnamefont {J.}~\bibnamefont {{Schwab}}},
  \bibinfo {author} {\bibfnamefont {E.}~\bibnamefont {{Bauer}}}, \bibinfo
  {author} {\bibfnamefont {F.~X.}\ \bibnamefont {{Timmes}}},\ and\ \bibinfo
  {author} {\bibfnamefont {A.~Y.}\ \bibnamefont {{Potekhin}}},\ }\bibfield
  {title} {\bibinfo {title} {{Skye: A Differentiable Equation of State}},\
  }\href {https://doi.org/10.3847/1538-4357/abf48e} {\bibfield  {journal}
  {\bibinfo  {journal} {\apj}\ }\textbf {\bibinfo {volume} {913}},\ \bibinfo
  {eid} {72} (\bibinfo {year} {2021})},\ \Eprint
  {https://arxiv.org/abs/2104.00691} {arXiv:2104.00691 [astro-ph.SR]}
  \BibitemShut {NoStop}%
\bibitem [{\citenamefont {{Iglesias}}\ and\ \citenamefont
  {{Rogers}}(1993)}]{Iglesias1993}%
  \BibitemOpen
  \bibfield  {author} {\bibinfo {author} {\bibfnamefont {C.~A.}\ \bibnamefont
  {{Iglesias}}}\ and\ \bibinfo {author} {\bibfnamefont {F.~J.}\ \bibnamefont
  {{Rogers}}},\ }\bibfield  {title} {\bibinfo {title} {{Radiative opacities for
  carbon- and oxygen-rich mixtures}},\ }\href {https://doi.org/10.1086/172958}
  {\bibfield  {journal} {\bibinfo  {journal} {\apj}\ }\textbf {\bibinfo
  {volume} {412}},\ \bibinfo {pages} {752} (\bibinfo {year}
  {1993})}\BibitemShut {NoStop}%
\bibitem [{\citenamefont {{Iglesias}}\ and\ \citenamefont
  {{Rogers}}(1996)}]{Iglesias1996}%
  \BibitemOpen
  \bibfield  {author} {\bibinfo {author} {\bibfnamefont {C.~A.}\ \bibnamefont
  {{Iglesias}}}\ and\ \bibinfo {author} {\bibfnamefont {F.~J.}\ \bibnamefont
  {{Rogers}}},\ }\bibfield  {title} {\bibinfo {title} {{Updated Opal
  Opacities}},\ }\href {https://doi.org/10.1086/177381} {\bibfield  {journal}
  {\bibinfo  {journal} {\apj}\ }\textbf {\bibinfo {volume} {464}},\ \bibinfo
  {pages} {943} (\bibinfo {year} {1996})}\BibitemShut {NoStop}%
\bibitem [{\citenamefont {{Ferguson}}\ \emph {et~al.}(2005)\citenamefont
  {{Ferguson}}, \citenamefont {{Alexander}}, \citenamefont {{Allard}},
  \citenamefont {{Barman}}, \citenamefont {{Bodnarik}}, \citenamefont
  {{Hauschildt}}, \citenamefont {{Heffner-Wong}},\ and\ \citenamefont
  {{Tamanai}}}]{Ferguson2005}%
  \BibitemOpen
  \bibfield  {author} {\bibinfo {author} {\bibfnamefont {J.~W.}\ \bibnamefont
  {{Ferguson}}}, \bibinfo {author} {\bibfnamefont {D.~R.}\ \bibnamefont
  {{Alexander}}}, \bibinfo {author} {\bibfnamefont {F.}~\bibnamefont
  {{Allard}}}, \bibinfo {author} {\bibfnamefont {T.}~\bibnamefont {{Barman}}},
  \bibinfo {author} {\bibfnamefont {J.~G.}\ \bibnamefont {{Bodnarik}}},
  \bibinfo {author} {\bibfnamefont {P.~H.}\ \bibnamefont {{Hauschildt}}},
  \bibinfo {author} {\bibfnamefont {A.}~\bibnamefont {{Heffner-Wong}}},\ and\
  \bibinfo {author} {\bibfnamefont {A.}~\bibnamefont {{Tamanai}}},\ }\bibfield
  {title} {\bibinfo {title} {{Low-Temperature Opacities}},\ }\href
  {https://doi.org/10.1086/428642} {\bibfield  {journal} {\bibinfo  {journal}
  {\apj}\ }\textbf {\bibinfo {volume} {623}},\ \bibinfo {pages} {585} (\bibinfo
  {year} {2005})},\ \Eprint {https://arxiv.org/abs/astro-ph/0502045}
  {astro-ph/0502045} \BibitemShut {NoStop}%
\bibitem [{\citenamefont {{Poutanen}}(2017)}]{Poutanen2017}%
  \BibitemOpen
  \bibfield  {author} {\bibinfo {author} {\bibfnamefont {J.}~\bibnamefont
  {{Poutanen}}},\ }\bibfield  {title} {\bibinfo {title} {{Rosseland and Flux
  Mean Opacities for Compton Scattering}},\ }\href
  {https://doi.org/10.3847/1538-4357/835/2/119} {\bibfield  {journal} {\bibinfo
   {journal} {\apj}\ }\textbf {\bibinfo {volume} {835}},\ \bibinfo {eid} {119}
  (\bibinfo {year} {2017})},\ \Eprint {https://arxiv.org/abs/1606.09466}
  {arXiv:1606.09466 [astro-ph.HE]} \BibitemShut {NoStop}%
\bibitem [{\citenamefont {{Cassisi}}\ \emph {et~al.}(2007)\citenamefont
  {{Cassisi}}, \citenamefont {{Potekhin}}, \citenamefont {{Pietrinferni}},
  \citenamefont {{Catelan}},\ and\ \citenamefont {{Salaris}}}]{Cassisi2007}%
  \BibitemOpen
  \bibfield  {author} {\bibinfo {author} {\bibfnamefont {S.}~\bibnamefont
  {{Cassisi}}}, \bibinfo {author} {\bibfnamefont {A.~Y.}\ \bibnamefont
  {{Potekhin}}}, \bibinfo {author} {\bibfnamefont {A.}~\bibnamefont
  {{Pietrinferni}}}, \bibinfo {author} {\bibfnamefont {M.}~\bibnamefont
  {{Catelan}}},\ and\ \bibinfo {author} {\bibfnamefont {M.}~\bibnamefont
  {{Salaris}}},\ }\bibfield  {title} {\bibinfo {title} {{Updated
  Electron-Conduction Opacities: The Impact on Low-Mass Stellar Models}},\
  }\href {https://doi.org/10.1086/516819} {\bibfield  {journal} {\bibinfo
  {journal} {\apj}\ }\textbf {\bibinfo {volume} {661}},\ \bibinfo {pages}
  {1094} (\bibinfo {year} {2007})},\ \Eprint
  {https://arxiv.org/abs/astro-ph/0703011} {astro-ph/0703011} \BibitemShut
  {NoStop}%
\bibitem [{\citenamefont {{Cyburt}}\ \emph {et~al.}(2010)\citenamefont
  {{Cyburt}}, \citenamefont {{Amthor}}, \citenamefont {{Ferguson}},
  \citenamefont {{Meisel}}, \citenamefont {{Smith}}, \citenamefont {{Warren}},
  \citenamefont {{Heger}}, \citenamefont {{Hoffman}}, \citenamefont
  {{Rauscher}}, \citenamefont {{Sakharuk}}, \citenamefont {{Schatz}},
  \citenamefont {{Thielemann}},\ and\ \citenamefont {{Wiescher}}}]{Cyburt2010}%
  \BibitemOpen
  \bibfield  {author} {\bibinfo {author} {\bibfnamefont {R.~H.}\ \bibnamefont
  {{Cyburt}}}, \bibinfo {author} {\bibfnamefont {A.~M.}\ \bibnamefont
  {{Amthor}}}, \bibinfo {author} {\bibfnamefont {R.}~\bibnamefont
  {{Ferguson}}}, \bibinfo {author} {\bibfnamefont {Z.}~\bibnamefont
  {{Meisel}}}, \bibinfo {author} {\bibfnamefont {K.}~\bibnamefont {{Smith}}},
  \bibinfo {author} {\bibfnamefont {S.}~\bibnamefont {{Warren}}}, \bibinfo
  {author} {\bibfnamefont {A.}~\bibnamefont {{Heger}}}, \bibinfo {author}
  {\bibfnamefont {R.~D.}\ \bibnamefont {{Hoffman}}}, \bibinfo {author}
  {\bibfnamefont {T.}~\bibnamefont {{Rauscher}}}, \bibinfo {author}
  {\bibfnamefont {A.}~\bibnamefont {{Sakharuk}}}, \bibinfo {author}
  {\bibfnamefont {H.}~\bibnamefont {{Schatz}}}, \bibinfo {author}
  {\bibfnamefont {F.~K.}\ \bibnamefont {{Thielemann}}},\ and\ \bibinfo {author}
  {\bibfnamefont {M.}~\bibnamefont {{Wiescher}}},\ }\bibfield  {title}
  {\bibinfo {title} {{The JINA REACLIB Database: Its Recent Updates and Impact
  on Type-I X-ray Bursts}},\ }\href
  {https://doi.org/10.1088/0067-0049/189/1/240} {\bibfield  {journal} {\bibinfo
   {journal} {{ApJ}}\ }\textbf {\bibinfo {volume} {189}},\ \bibinfo {pages}
  {240} (\bibinfo {year} {2010})}\BibitemShut {NoStop}%
\bibitem [{\citenamefont {{Angulo}}\ \emph {et~al.}(1999)\citenamefont
  {{Angulo}}, \citenamefont {{Arnould}}, \citenamefont {{Rayet}}, \citenamefont
  {{Descouvemont}}, \citenamefont {{Baye}}, \citenamefont {{Leclercq-Willain}},
  \citenamefont {{Coc}}, \citenamefont {{Barhoumi}}, \citenamefont {{Aguer}},
  \citenamefont {{Rolfs}}, \citenamefont {{Kunz}}, \citenamefont {{Hammer}},
  \citenamefont {{Mayer}}, \citenamefont {{Paradellis}}, \citenamefont
  {{Kossionides}}, \citenamefont {{Chronidou}}, \citenamefont {{Spyrou}},
  \citenamefont {{degl'Innocenti}}, \citenamefont {{Fiorentini}}, \citenamefont
  {{Ricci}}, \citenamefont {{Zavatarelli}}, \citenamefont {{Providencia}},
  \citenamefont {{Wolters}}, \citenamefont {{Soares}}, \citenamefont {{Grama}},
  \citenamefont {{Rahighi}}, \citenamefont {{Shotter}},\ and\ \citenamefont
  {{Lamehi Rachti}}}]{Angulo1999}%
  \BibitemOpen
  \bibfield  {author} {\bibinfo {author} {\bibfnamefont {C.}~\bibnamefont
  {{Angulo}}}, \bibinfo {author} {\bibfnamefont {M.}~\bibnamefont {{Arnould}}},
  \bibinfo {author} {\bibfnamefont {M.}~\bibnamefont {{Rayet}}}, \bibinfo
  {author} {\bibfnamefont {P.}~\bibnamefont {{Descouvemont}}}, \bibinfo
  {author} {\bibfnamefont {D.}~\bibnamefont {{Baye}}}, \bibinfo {author}
  {\bibfnamefont {C.}~\bibnamefont {{Leclercq-Willain}}}, \bibinfo {author}
  {\bibfnamefont {A.}~\bibnamefont {{Coc}}}, \bibinfo {author} {\bibfnamefont
  {S.}~\bibnamefont {{Barhoumi}}}, \bibinfo {author} {\bibfnamefont
  {P.}~\bibnamefont {{Aguer}}}, \bibinfo {author} {\bibfnamefont
  {C.}~\bibnamefont {{Rolfs}}}, \bibinfo {author} {\bibfnamefont
  {R.}~\bibnamefont {{Kunz}}}, \bibinfo {author} {\bibfnamefont {J.~W.}\
  \bibnamefont {{Hammer}}}, \bibinfo {author} {\bibfnamefont {A.}~\bibnamefont
  {{Mayer}}}, \bibinfo {author} {\bibfnamefont {T.}~\bibnamefont
  {{Paradellis}}}, \bibinfo {author} {\bibfnamefont {S.}~\bibnamefont
  {{Kossionides}}}, \bibinfo {author} {\bibfnamefont {C.}~\bibnamefont
  {{Chronidou}}}, \bibinfo {author} {\bibfnamefont {K.}~\bibnamefont
  {{Spyrou}}}, \bibinfo {author} {\bibfnamefont {S.}~\bibnamefont
  {{degl'Innocenti}}}, \bibinfo {author} {\bibfnamefont {G.}~\bibnamefont
  {{Fiorentini}}}, \bibinfo {author} {\bibfnamefont {B.}~\bibnamefont
  {{Ricci}}}, \bibinfo {author} {\bibfnamefont {S.}~\bibnamefont
  {{Zavatarelli}}}, \bibinfo {author} {\bibfnamefont {C.}~\bibnamefont
  {{Providencia}}}, \bibinfo {author} {\bibfnamefont {H.}~\bibnamefont
  {{Wolters}}}, \bibinfo {author} {\bibfnamefont {J.}~\bibnamefont {{Soares}}},
  \bibinfo {author} {\bibfnamefont {C.}~\bibnamefont {{Grama}}}, \bibinfo
  {author} {\bibfnamefont {J.}~\bibnamefont {{Rahighi}}}, \bibinfo {author}
  {\bibfnamefont {A.}~\bibnamefont {{Shotter}}},\ and\ \bibinfo {author}
  {\bibfnamefont {M.}~\bibnamefont {{Lamehi Rachti}}},\ }\bibfield  {title}
  {\bibinfo {title} {{A compilation of charged-particle induced thermonuclear
  reaction rates}},\ }\href {https://doi.org/10.1016/S0375-9474(99)00030-5}
  {\bibfield  {journal} {\bibinfo  {journal} {Nuclear Physics A}\ }\textbf
  {\bibinfo {volume} {656}},\ \bibinfo {pages} {3} (\bibinfo {year}
  {1999})}\BibitemShut {NoStop}%
\bibitem [{\citenamefont {{Fuller}}\ \emph {et~al.}(1985)\citenamefont
  {{Fuller}}, \citenamefont {{Fowler}},\ and\ \citenamefont
  {{Newman}}}]{Fuller1985}%
  \BibitemOpen
  \bibfield  {author} {\bibinfo {author} {\bibfnamefont {G.~M.}\ \bibnamefont
  {{Fuller}}}, \bibinfo {author} {\bibfnamefont {W.~A.}\ \bibnamefont
  {{Fowler}}},\ and\ \bibinfo {author} {\bibfnamefont {M.~J.}\ \bibnamefont
  {{Newman}}},\ }\bibfield  {title} {\bibinfo {title} {{Stellar weak
  interaction rates for intermediate-mass nuclei. IV - Interpolation procedures
  for rapidly varying lepton capture rates using effective log (ft)-values}},\
  }\href {https://doi.org/10.1086/163208} {\bibfield  {journal} {\bibinfo
  {journal} {\apj}\ }\textbf {\bibinfo {volume} {293}},\ \bibinfo {pages} {1}
  (\bibinfo {year} {1985})}\BibitemShut {NoStop}%
\bibitem [{\citenamefont {{Oda}}\ \emph {et~al.}(1994)\citenamefont {{Oda}},
  \citenamefont {{Hino}}, \citenamefont {{Muto}}, \citenamefont {{Takahara}},\
  and\ \citenamefont {{Sato}}}]{Oda1994}%
  \BibitemOpen
  \bibfield  {author} {\bibinfo {author} {\bibfnamefont {T.}~\bibnamefont
  {{Oda}}}, \bibinfo {author} {\bibfnamefont {M.}~\bibnamefont {{Hino}}},
  \bibinfo {author} {\bibfnamefont {K.}~\bibnamefont {{Muto}}}, \bibinfo
  {author} {\bibfnamefont {M.}~\bibnamefont {{Takahara}}},\ and\ \bibinfo
  {author} {\bibfnamefont {K.}~\bibnamefont {{Sato}}},\ }\bibfield  {title}
  {\bibinfo {title} {{Rate Tables for the Weak Processes of sd-Shell Nuclei in
  Stellar Matter}},\ }\href {https://doi.org/10.1006/adnd.1994.1007} {\bibfield
   {journal} {\bibinfo  {journal} {Atomic Data and Nuclear Data Tables}\
  }\textbf {\bibinfo {volume} {56}},\ \bibinfo {pages} {231} (\bibinfo {year}
  {1994})}\BibitemShut {NoStop}%
\bibitem [{\citenamefont {{Langanke}}\ and\ \citenamefont
  {{Mart{\'{\i}}nez-Pinedo}}(2000)}]{Langanke2000}%
  \BibitemOpen
  \bibfield  {author} {\bibinfo {author} {\bibfnamefont {K.}~\bibnamefont
  {{Langanke}}}\ and\ \bibinfo {author} {\bibfnamefont {G.}~\bibnamefont
  {{Mart{\'{\i}}nez-Pinedo}}},\ }\bibfield  {title} {\bibinfo {title}
  {{Shell-model calculations of stellar weak interaction rates: II. Weak rates
  for nuclei in the mass range /A=45-65 in supernovae environments}},\ }\href
  {https://doi.org/10.1016/S0375-9474(00)00131-7} {\bibfield  {journal}
  {\bibinfo  {journal} {Nuclear Physics A}\ }\textbf {\bibinfo {volume}
  {673}},\ \bibinfo {pages} {481} (\bibinfo {year} {2000})},\ \Eprint
  {https://arxiv.org/abs/nucl-th/0001018} {nucl-th/0001018} \BibitemShut
  {NoStop}%
\bibitem [{\citenamefont {{Chugunov}}\ \emph {et~al.}(2007)\citenamefont
  {{Chugunov}}, \citenamefont {{Dewitt}},\ and\ \citenamefont
  {{Yakovlev}}}]{Chugunov2007}%
  \BibitemOpen
  \bibfield  {author} {\bibinfo {author} {\bibfnamefont {A.~I.}\ \bibnamefont
  {{Chugunov}}}, \bibinfo {author} {\bibfnamefont {H.~E.}\ \bibnamefont
  {{Dewitt}}},\ and\ \bibinfo {author} {\bibfnamefont {D.~G.}\ \bibnamefont
  {{Yakovlev}}},\ }\bibfield  {title} {\bibinfo {title} {{Coulomb tunneling for
  fusion reactions in dense matter: Path integral MonteCarlo versus mean
  field}},\ }\href {https://doi.org/10.1103/PhysRevD.76.025028} {\bibfield
  {journal} {\bibinfo  {journal} {\prd}\ }\textbf {\bibinfo {volume} {76}},\
  \bibinfo {eid} {025028} (\bibinfo {year} {2007})},\ \Eprint
  {https://arxiv.org/abs/0707.3500} {arXiv:0707.3500} \BibitemShut {NoStop}%
\bibitem [{\citenamefont {{Itoh}}\ \emph {et~al.}(1996)\citenamefont {{Itoh}},
  \citenamefont {{Hayashi}}, \citenamefont {{Nishikawa}},\ and\ \citenamefont
  {{Kohyama}}}]{Itoh1996}%
  \BibitemOpen
  \bibfield  {author} {\bibinfo {author} {\bibfnamefont {N.}~\bibnamefont
  {{Itoh}}}, \bibinfo {author} {\bibfnamefont {H.}~\bibnamefont {{Hayashi}}},
  \bibinfo {author} {\bibfnamefont {A.}~\bibnamefont {{Nishikawa}}},\ and\
  \bibinfo {author} {\bibfnamefont {Y.}~\bibnamefont {{Kohyama}}},\ }\bibfield
  {title} {\bibinfo {title} {{Neutrino Energy Loss in Stellar Interiors. VII.
  Pair, Photo-, Plasma, Bremsstrahlung, and Recombination Neutrino
  Processes}},\ }\href {https://doi.org/10.1086/192264} {\bibfield  {journal}
  {\bibinfo  {journal} {{ApJ}}\ }\textbf {\bibinfo {volume} {102}},\ \bibinfo
  {pages} {411} (\bibinfo {year} {1996})}\BibitemShut {NoStop}%
\bibitem [{\citenamefont {{Cox}}\ and\ \citenamefont
  {{Giuli}}(1968)}]{1968pss..book.....C}%
  \BibitemOpen
  \bibfield  {author} {\bibinfo {author} {\bibfnamefont {J.~P.}\ \bibnamefont
  {{Cox}}}\ and\ \bibinfo {author} {\bibfnamefont {R.~T.}\ \bibnamefont
  {{Giuli}}},\ }\href@noop {} {\emph {\bibinfo {title} {{Principles of stellar
  structure}}}}\ (\bibinfo {year} {1968})\BibitemShut {NoStop}%
\bibitem [{\citenamefont {Bahcall}\ \emph {et~al.}(2006)\citenamefont
  {Bahcall}, \citenamefont {Serenelli},\ and\ \citenamefont
  {Basu}}]{Bahcall:2005va}%
  \BibitemOpen
  \bibfield  {author} {\bibinfo {author} {\bibfnamefont {J.~N.}\ \bibnamefont
  {Bahcall}}, \bibinfo {author} {\bibfnamefont {A.~M.}\ \bibnamefont
  {Serenelli}},\ and\ \bibinfo {author} {\bibfnamefont {S.}~\bibnamefont
  {Basu}},\ }\bibfield  {title} {\bibinfo {title} {{10,000 standard solar
  models: a Monte Carlo simulation}},\ }\href {https://doi.org/10.1086/504043}
  {\bibfield  {journal} {\bibinfo  {journal} {Astrophys. J. Suppl.}\ }\textbf
  {\bibinfo {volume} {165}},\ \bibinfo {pages} {400} (\bibinfo {year}
  {2006})},\ \Eprint {https://arxiv.org/abs/astro-ph/0511337}
  {arXiv:astro-ph/0511337} \BibitemShut {NoStop}%
\bibitem [{\citenamefont {Grevesse}\ and\ \citenamefont
  {Sauval}(1998)}]{Grevesse:1998bj}%
  \BibitemOpen
  \bibfield  {author} {\bibinfo {author} {\bibfnamefont {N.}~\bibnamefont
  {Grevesse}}\ and\ \bibinfo {author} {\bibfnamefont {A.~J.}\ \bibnamefont
  {Sauval}},\ }\bibfield  {title} {\bibinfo {title} {{Standard Solar
  Composition}},\ }\href {https://doi.org/10.1023/A:1005161325181} {\bibfield
  {journal} {\bibinfo  {journal} {Space Sci. Rev.}\ }\textbf {\bibinfo {volume}
  {85}},\ \bibinfo {pages} {161} (\bibinfo {year} {1998})}\BibitemShut
  {NoStop}%
\bibitem [{\citenamefont {Metcalfe}\ \emph {et~al.}(2012)\citenamefont
  {Metcalfe} \emph {et~al.}}]{Metcalfe:2012mu}%
  \BibitemOpen
  \bibfield  {author} {\bibinfo {author} {\bibfnamefont {T.~S.}\ \bibnamefont
  {Metcalfe}} \emph {et~al.},\ }\bibfield  {title} {\bibinfo {title}
  {{Asteroseismology of the solar analogs 16 Cyg A \& B from Kepler
  observations}},\ }\href {https://doi.org/10.1088/2041-8205/748/1/L10}
  {\bibfield  {journal} {\bibinfo  {journal} {Astrophys. J. Lett.}\ }\textbf
  {\bibinfo {volume} {748}},\ \bibinfo {pages} {L10} (\bibinfo {year}
  {2012})},\ \Eprint {https://arxiv.org/abs/1201.5966} {arXiv:1201.5966
  [astro-ph.SR]} \BibitemShut {NoStop}%
\bibitem [{\citenamefont {Capelo}\ and\ \citenamefont
  {Lopes}(2020)}]{Capelo:2020lha}%
  \BibitemOpen
  \bibfield  {author} {\bibinfo {author} {\bibfnamefont {D.}~\bibnamefont
  {Capelo}}\ and\ \bibinfo {author} {\bibfnamefont {I.}~\bibnamefont {Lopes}},\
  }\bibfield  {title} {\bibinfo {title} {{The impact of composition choices on
  solar evolution: age, helio- and asteroseismology, and neutrinos}},\ }\href
  {https://doi.org/10.1093/mnras/staa2402} {\bibfield  {journal} {\bibinfo
  {journal} {Mon. Not. Roy. Astron. Soc.}\ }\textbf {\bibinfo {volume} {498}},\
  \bibinfo {pages} {1992} (\bibinfo {year} {2020})},\ \Eprint
  {https://arxiv.org/abs/2010.01686} {arXiv:2010.01686 [astro-ph.SR]}
  \BibitemShut {NoStop}%
\bibitem [{\citenamefont {Nelder}\ and\ \citenamefont
  {Mead}(1965)}]{Nelder:1965zz}%
  \BibitemOpen
  \bibfield  {author} {\bibinfo {author} {\bibfnamefont {J.~A.}\ \bibnamefont
  {Nelder}}\ and\ \bibinfo {author} {\bibfnamefont {R.}~\bibnamefont {Mead}},\
  }\bibfield  {title} {\bibinfo {title} {{A Simplex Method for Function
  Minimization}},\ }\href {https://doi.org/10.1093/comjnl/7.4.308} {\bibfield
  {journal} {\bibinfo  {journal} {Comput. J.}\ }\textbf {\bibinfo {volume}
  {7}},\ \bibinfo {pages} {308} (\bibinfo {year} {1965})}\BibitemShut {NoStop}%
\bibitem [{\citenamefont {Hekker}\ and\ \citenamefont
  {Mazumdar}(2013)}]{hekker_mazumdar_2013}%
  \BibitemOpen
  \bibfield  {author} {\bibinfo {author} {\bibfnamefont {S.}~\bibnamefont
  {Hekker}}\ and\ \bibinfo {author} {\bibfnamefont {A.}~\bibnamefont
  {Mazumdar}},\ }\bibfield  {title} {\bibinfo {title} {Solar-like oscillations
  in subgiant and red-giant stars: mixed modes},\ }\href
  {https://doi.org/10.1017/S1743921313014531} {\bibfield  {journal} {\bibinfo
  {journal} {Proceedings of the International Astronomical Union}\ }\textbf
  {\bibinfo {volume} {9}},\ \bibinfo {pages} {325} (\bibinfo {year}
  {2013})}\BibitemShut {NoStop}%
\bibitem [{\citenamefont {Kippenhahn}\ \emph {et~al.}(2012)\citenamefont
  {Kippenhahn}, \citenamefont {Weigert},\ and\ \citenamefont
  {Weiss}}]{Kippenhahn:2012qhp}%
  \BibitemOpen
  \bibfield  {author} {\bibinfo {author} {\bibfnamefont {R.}~\bibnamefont
  {Kippenhahn}}, \bibinfo {author} {\bibfnamefont {A.}~\bibnamefont
  {Weigert}},\ and\ \bibinfo {author} {\bibfnamefont {A.}~\bibnamefont
  {Weiss}},\ }\href {https://doi.org/10.1007/978-3-642-30304-3} {\emph
  {\bibinfo {title} {{Stellar structure and evolution}}}},\ Vol.\ \bibinfo
  {volume} {9783642303043}\ (\bibinfo  {publisher} {Springer},\ \bibinfo {year}
  {2012})\BibitemShut {NoStop}%
\bibitem [{\citenamefont {Lamers}\ and\ \citenamefont
  {M.~Levesque}(2017)}]{10.1088/978-0-7503-1278-3ch14}%
  \BibitemOpen
  \bibfield  {author} {\bibinfo {author} {\bibfnamefont {H.~J.}\ \bibnamefont
  {Lamers}}\ and\ \bibinfo {author} {\bibfnamefont {E.}~\bibnamefont
  {M.~Levesque}},\ }\bibfield  {title} {\bibinfo {title} {Principles of
  post-main-sequence evolution},\ }in\ \href
  {https://doi.org/10.1088/978-0-7503-1278-3ch14} {\emph {\bibinfo {booktitle}
  {Understanding Stellar Evolution}}},\ \bibinfo {series and number}
  {2514-3433}\ (\bibinfo  {publisher} {IOP Publishing},\ \bibinfo {year}
  {2017})\ pp.\ \bibinfo {pages} {14--1 to 14--10}\BibitemShut {NoStop}%
\bibitem [{\citenamefont {Townsend}\ and\ \citenamefont
  {Teitler}(2013)}]{Townsend:2013lua}%
  \BibitemOpen
  \bibfield  {author} {\bibinfo {author} {\bibfnamefont {R.~H.~D.}\
  \bibnamefont {Townsend}}\ and\ \bibinfo {author} {\bibfnamefont
  {S.}~\bibnamefont {Teitler}},\ }\bibfield  {title} {\bibinfo {title} {{GYRE:
  An open-source stellar oscillation code based on a new Magnus Multiple
  Shooting Scheme}},\ }\href {https://doi.org/10.1093/mnras/stt1533} {\bibfield
   {journal} {\bibinfo  {journal} {Mon. Not. Roy. Astron. Soc.}\ }\textbf
  {\bibinfo {volume} {435}},\ \bibinfo {pages} {3406} (\bibinfo {year}
  {2013})},\ \Eprint {https://arxiv.org/abs/1308.2965} {arXiv:1308.2965
  [astro-ph.SR]} \BibitemShut {NoStop}%
\bibitem [{\citenamefont {Irastorza}\ \emph {et~al.}(2012)\citenamefont
  {Irastorza} \emph {et~al.}}]{iaxo}%
  \BibitemOpen
  \bibfield  {author} {\bibinfo {author} {\bibfnamefont {I.~G.}\ \bibnamefont
  {Irastorza}} \emph {et~al.} (\bibinfo {collaboration} {IAXO}),\ }\bibfield
  {title} {\bibinfo {title} {{The International Axion Observatory (IAXO)}},\
  }in\ \href {https://doi.org/10.3204/DESY-PROC-2011-04/irastorza_igor} {\emph
  {\bibinfo {booktitle} {{7th Patras Workshop on Axions, WIMPs and WISPs}}}}\
  (\bibinfo {year} {2012})\ pp.\ \bibinfo {pages} {98--101},\ \Eprint
  {https://arxiv.org/abs/1201.3849} {arXiv:1201.3849 [hep-ex]} \BibitemShut
  {NoStop}%
\bibitem [{\citenamefont {Friedland}\ \emph
  {et~al.}(2013{\natexlab{b}})\citenamefont {Friedland}, \citenamefont
  {Giannotti},\ and\ \citenamefont {Wise}}]{Friedland:2012hj}%
  \BibitemOpen
  \bibfield  {author} {\bibinfo {author} {\bibfnamefont {A.}~\bibnamefont
  {Friedland}}, \bibinfo {author} {\bibfnamefont {M.}~\bibnamefont
  {Giannotti}},\ and\ \bibinfo {author} {\bibfnamefont {M.}~\bibnamefont
  {Wise}},\ }\bibfield  {title} {\bibinfo {title} {{Constraining the
  Axion-Photon Coupling with Massive Stars}},\ }\href
  {https://doi.org/10.1103/PhysRevLett.110.061101} {\bibfield  {journal}
  {\bibinfo  {journal} {Phys. Rev. Lett.}\ }\textbf {\bibinfo {volume} {110}},\
  \bibinfo {pages} {061101} (\bibinfo {year} {2013}{\natexlab{b}})},\ \Eprint
  {https://arxiv.org/abs/1210.1271} {arXiv:1210.1271 [hep-ph]} \BibitemShut
  {NoStop}%
\end{thebibliography}%

\end{document}